\documentclass[aps,prl,twocolumn,groupedaddress]{revtex4-2}
\pdfoutput=1

\usepackage[latin1]{inputenc}
\usepackage{latexsym,diagbox,stackrel}
\usepackage{relsize}
\usepackage{shuffle}
\usepackage{amsmath}
\usepackage{mathrsfs}
\usepackage{array}
\usepackage{arydshln,leftidx,mathtools}
\usepackage{bbold}
\usepackage{graphicx}
\usepackage{subcaption}
\usepackage{amssymb}
\usepackage{multirow}
\usepackage{enumerate}
\usepackage{blkarray}
\usepackage[usenames,dvipsnames,svgnames,table]{xcolor}
\usepackage[hidelinks]{hyperref}
\usepackage{cleveref}
\usepackage{slashed}
\usepackage{datetime}
\usepackage{tikz}
\usepackage{graphicx}
\usetikzlibrary{shapes.geometric, arrows}
\definecolor{light-gray}{gray}{0.95}
\definecolor{light-grayII}{gray}{0.85}
\usepackage{array}
\usepackage[colorinlistoftodos]{todonotes}

\begin{document}

\def\d{\text{d}}
\def\bd {\boldsymbol}
\newcommand{\C}{\mathbb{C}}
\newcommand{\CP}{\mathbb{CP}}
\newcommand{\PT}{\mathbb{PT}}
\newcommand{\R}{\mathbb{R}}
\newcommand{\bbS}{\mathbb{S}}
\newcommand{\cH}{\mathcal{H}}
\newcommand{\cI}{\mathcal{I}}
\newcommand{\cC}{\mathcal{C}}
\newcommand{\scri}{\mathscr{I}}
\newcommand{\cPT}{\mathcal{PT}}
\newcommand{\M}{\mathbb{M}}
\newcommand{\CM}{\mathscr{M}}
\newcommand{\tCM}{\widetilde{\mathscr{M}}}
\newcommand{\T}{\mathbb{T}}
\newcommand{\Z}{\mathbb{Z}}
\newcommand{\p}{\partial}
\newcommand{\dbar}{\bar\partial}
\newcommand{\g}{\mathfrak{g}}
\newcommand{\cM}{\mathcal{M}}
\newcommand{\SL}{\mathrm{SL}}
\newcommand{\End}{\mathrm{End}}
\newcommand{\tr}{\mathrm{tr}}
\newcommand{\sgn}{\mathrm{sgn}}
\newcommand{\SU}{\, \mathrm{SU}}
\newcommand{\GL}{\mathrm{GL}}
\newcommand{\diag}{\, \mathrm{diag}}
\newcommand{\rd}{\, \mathrm{d}}
\newcommand{\pf}{\text{Pf} \,}
\newcommand{\pfr}{\text{Pf}^\prime \,}
\newcommand{\precs}[1]{\prec_{\!\mathsmaller{#1}}}
\newcommand{\px}{\raisebox{.15\baselineskip}{\Large\ensuremath{\wp}}}
\newcommand{\vol}{\mathrm{vol}\,}
\newcommand{\yhat}{\hat{\mathcal{Y}}}
\newcommand{\trho}{\tilde{\rho}}
\newcommand{\sL}{{\scalebox{0.6}{$L$}}}
\newcommand{\sR}{{\scalebox{0.6}{$R$}}}
\newcommand{\ssL}{{\scalebox{0.5}{$L$}}}
\newcommand{\ssR}{{\scalebox{0.5}{$R$}}}
\newcommand{\onep}{L_1^+}
\newcommand{\onem}{L_1^-}
\newcommand{\twop}{L_2^+}
\newcommand{\twom}{L_2^-}
\newcommand{\D}[2]{\delta_{#1,#2}}

\newcommand{\addcite}{\todo[color=OrangeRed]{Add citation.}}
\newcommand{\rewrite}[1]{\todo[color=green!40]{#1}}
\newcommand{\todoRS}[1]{\todo[color=Dandelion]{#1}}
\newcommand{\todoRM}[1]{\todo[color=Cerulean]{#1}}
\newcommand{\todoY}[1]{\todo[color=LimeGreen]{#1}}




\begin{flushright}
\hspace{90mm} QMUL-PH-21-29
\end{flushright}
 
\title{\Large Superstring loop amplitudes from the field theory limit}

%
%
%

\author{Yvonne Geyer}
\email[]{yjgeyer@gmail.com}
\affiliation{Department of Physics, Faculty of Science, Chulalongkorn University\\
Thanon Phayathai, Pathumwan, Bangkok 10330, Thailand}

\author{Ricardo Monteiro}
\email[]{ricardo.monteiro@qmul.ac.uk}
\affiliation{Centre for Research in String Theory, School of Physics and Astronomy \\
Queen Mary University of London, E1 4NS, United Kingdom}
        
\author{Ricardo Stark-Much\~ao}
\email[]{r.j.stark-muchao@qmul.ac.uk}
\affiliation{Centre for Research in String Theory, School of Physics and Astronomy \\
Queen Mary University of London, E1 4NS, United Kingdom}

\begin{abstract}
We propose a procedure to determine the moduli-space integrands of loop-level superstring amplitudes for massless external states in terms of the field theory limit. We focus on the type II superstring. The procedure is to: (i) take a supergravity loop integrand written in a BCJ double-copy representation, (ii) use the loop-level scattering equations to translate that integrand into the ambitwistor string moduli-space integrand, localised on the nodal Riemann sphere, and (iii) uplift that formula to one on the higher-genus surface valid for the superstring, guided by modular invariance. We show how this works for the four-point amplitude at two loops, where we reproduce the known answer, and at three loops, where we present a conjecture that is consistent with a previous proposal for the chiral measure. Useful supergravity results are currently known up to five loops.
\end{abstract}

\maketitle

\section{Introduction}

The birth of string theory is widely considered to be the discovery by Veneziano of the scattering amplitude formula that today bears his name \cite{Veneziano:1968yb}. More than five decades later, the calculation of string scattering amplitudes remains a formidable challenge. To give the example of the type II superstring in Minkowski spacetime, the four-point amplitude for massless external states was computed at tree level and one loop in 1982 \cite{Green:1981yb,Schwarz:1982jn}, and at two loops in 2005 \cite{DHoker:2005vch,Berkovits:2005df,Berkovits:2005ng}. There has been significant work on the three-loop problem, namely a proposal for the chiral measure \cite{DHoker:2004fcs,DHoker:2004qhf,Cacciatori:2008ay} and a partial computation using the pure spinor formalism \cite{Gomez:2013sla}, but it remains to be fully addressed. The advances have had a rich interplay with those in gauge theory and gravity amplitudes, particularly in their maximally supersymmetric versions. For instance, the first computations of the four-point one-loop amplitudes in the now widely studied 4D ${\mathcal N}=4$ super-Yang-Mills theory (SYM) and ${\mathcal N}=8$ supergravity were based on the field theory limit of the analogous superstring calculations \cite{Green:1982sw}. In this paper, we aim to return the favour by importing three-loop results in ${\mathcal N}=8$ supergravity, themselves obtained from non-planar ${\mathcal N}=4$ SYM via the Bern-Carrasco-Johansson (BCJ) double copy \cite{Bern:2010ue}, into the type II superstring.

\section{String theory versus field theory}

We will consider the type II superstring four-point amplitude for massless incoming states of momenta $k_i$ $(i=1,\ldots,4)$. The 10D maximal supersymmetry implies that information on the four external states is encoded in a kinematic prefactor ${\mathcal R}^4$ \cite{Green:1987sp}, such that the supergravity tree-level amplitude is $\sim{\mathcal R}^4/(s_{12}s_{13}s_{14})$. We define the Mandelstam variables as $s_{ij}=2k_i\cdot k_j$. Our working assumption will be that, up to three loops \footnote{Beyond three loops, the integration over the period matrix $\Omega_{IJ}$ must be restricted due to the Schottky problem. Beyond four loops, the delicate issue of non-projectedness of supermoduli space is also known to arise \cite{Donagi:2013dua}.}, the $g$-loop superstring amplitude ${\mathcal A}_{\,\mathbb{S}}^{(g)}$ takes the form
\begin{align}
\label{eq:ssamp}
& \frac{{\mathcal A}_{\,\mathbb{S}}^{(g)}}{{\mathcal R}^4} =  \int_{{\mathcal M}_{g,4}} \Big|\! \prod_{I\leq J} d\Omega_{IJ\,} \! \Big|^2 \int \!\! d\ell \,\, \big|{\mathcal Y}_{\mathbb{S}}^{(g)}\big|^2 \,\prod_{i<j} | E(z_i,z_j) |^{\frac{\alpha'  \!s_{ij}}{2}} \nonumber\\
& \; \times 
\Big|\exp{\frac{\alpha'}{2}\!\big(i\pi\, \Omega_{IJ}\,\ell^I\!\cdot\! \ell^J+2\pi i\sum_j \ell^I \!\cdot\! k_j\!\int_{z_0}^{z_j}\!\omega_I\big)}\Big|^{\,2} \,.
\end{align}
The integration denoted by ${{\mathcal M}_{g,4}}$ is over a genus-$g$ fundamental domain parametrised by the period matrix $\Omega_{IJ}$ ($I,J=1,\ldots,g$) and over four marked points $z_i$. We use a `chiral splitting' representation \cite{DHoker:1988ta,DHoker:1989cxq}, made possible by the introduction of the loop momenta $\ell^I$, with $d\ell$ denoting $\prod_I d^{10}\ell^I$. 
The prime form $E(z_i,z_j)$ and the exponential
(involving the normalised holomorphic Abelian differentials
$\omega_I$ whose $B$-periods define the period matrix)
constitute the chiral$\times$anti-chiral loop-level
Koba-Nielsen factors.
The interesting object is ${\mathcal Y}_{\,\mathbb{S}}^{(g)}$. We make no distinction between type IIA and type IIB apart from the details of ${\mathcal R}^4$, since at four points there is no contribution from odd spin structures at least up to three loops  \footnote{This follows from supersymmetry and is clear if we consider factorisable NS-NS external states $\varepsilon_i^{\mu\nu}=\epsilon_i^{\mu}\tilde\epsilon_i^{\nu}$. The supersymmetric pre-factor is ${\mathcal R}^4(\epsilon,\tilde\epsilon)={\mathcal F}^4(\epsilon){\mathcal F}^4(\tilde\epsilon)$, where ${\mathcal F}^4$ is the pre-factor for the open superstring and includes products $\epsilon_i\cdot\epsilon_j$. At three loops and four points, a 10D Levi-Civita tensor arising from an odd spin structure's zero mode may just about be saturated: $\varepsilon_{10}(k_1,k_2,k_3,\epsilon_1,\epsilon_2,\epsilon_3,\epsilon_4,\ell_1,\ell_2,\ell_3)$, but it would never give rise to any $\epsilon_i\cdot\epsilon_j$. Moreover, the contraction of the two Levi-Civita tensors (with $\epsilon_i$ and with $\tilde\epsilon_i$) over three indices after loop integration, required for a potentially non-vanishing contribution, yields products $\epsilon_i\cdot\tilde\epsilon_j$, inconsistent with ${\mathcal F}^4(\epsilon){\mathcal F}^4(\tilde\epsilon)$. This discussion is consistent with the results of \cite{Gomez:2013sla,Berkovits:2006vc}.}. 

We will exploit the analogy between the formula \eqref{eq:ssamp} for the superstring and the following expected formula for supergravity:
\begin{align}
\label{eq:asamp}
 \frac{\mathcal{A}_{\,\mathbb{A}}^{(g)}}{\mathcal{R}^4}
 & = \! \int \! d\ell \! \int_{{\mathcal M}_{g,4}}  \prod_{I\leq J} d\Omega_{IJ} \;\big(\mathcal{Y}_{\mathbb{A}}^{(g)}\big)^2  \prod_{i=1}^4\bar\delta(\mathcal{E}_i)\prod_{I\leq J}\bar\delta(u^{IJ})   \,.
\end{align}
This type of formula for a scattering amplitude was discovered at tree level by Cachazo, He and Yuan \cite{Cachazo:2013hca,Cachazo:2013iea}, generalising a previous formula from twistor string theory \cite{Witten:2003nn,Roiban:2004yf}.
The loop-level extension \cite{Adamo:2013tsa,Casali:2014hfa,Adamo:2015hoa,Geyer:2015bja,Geyer:2015jch,Geyer:2016wjx,Geyer:2018xwu} was derived from the type II ambitwistor string \cite{Mason:2013sva}, which is a worldsheet model of type II supergravity. The 10D loop integration in \eqref{eq:asamp} is UV divergent, so the expression is formal only, and we understand it as defining a loop integrand. The genus-$g$ moduli-space integration is fully localised on a set of critical points, determined by the genus-$g$ {\it scattering equations}: $\mathcal{E}_i=0$ and $u^{IJ}=0$ \footnote{For the delta functions, we use the definition $\bar\delta(z)=\bar\partial(1/2\pi i z)$, standard in this context}.  An extensive discussion of the loop-level version of this formalism was presented in \cite{Geyer:2018xwu}; the brief discussion below will be sufficient for our purposes. There is a clear analogy between \eqref{eq:ssamp} and  \eqref{eq:asamp}. Our proposal, under conditions to be discussed, is to identify the `chiral half-integrands',
\begin{equation}
\label{eq:YsYa}
 \mathcal{Y}^{(g)}_{\mathbb{S}}=\mathcal{Y}^{(g)}_{\mathbb{A}}\,,
\end{equation}
which is known to be possible for $g\leq 2$. Notice that \eqref{eq:ssamp} is a simplified expression where $ \mathcal{Y}^{(g)}_{\mathbb{S}}$ is independent of $\alpha'$. The idea is that we can import an ambitwistor string---i.e.~supergravity---result into the superstring.

\begin{figure}[t]
    \includegraphics[width=0.2\textwidth]{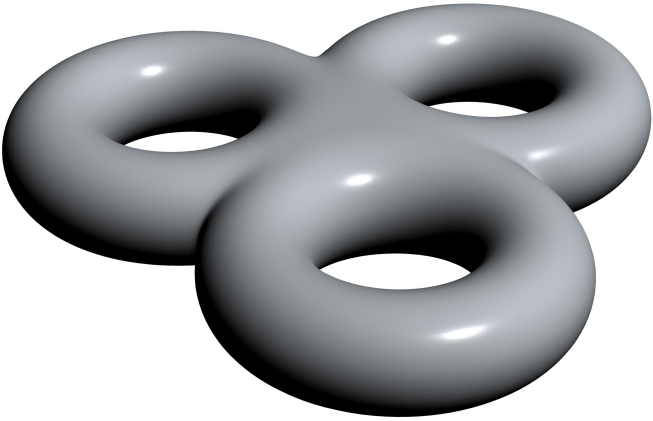} \qquad
    \includegraphics[width=0.2\textwidth]{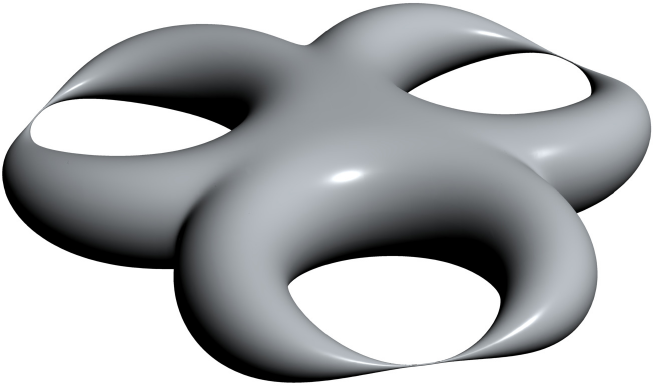}
\caption{Genus-3 surface and its maximal non-separating degeneration (genus 0) with 2 marked points per node.}
\label{fig:degen}
\end{figure}

The only known procedure to evaluate \eqref{eq:asamp} reflects the fact that the ambitwistor string is a field theory in disguise: the genus-$g$ formula can be localised on a maximal non-separating degeneration, i.e.~a Riemann sphere with $g$ nodes, as in FIG.~\ref{fig:degen}. This follows from a residue argument in moduli space at one \cite{Geyer:2015bja,Geyer:2015jch} and two \cite{Geyer:2016wjx,Geyer:2018xwu} loops, and our three-loop results provide evidence that it holds at higher order. The formula on the nodal sphere is
\begin{align}
\label{eq:assamp}
 \frac{\mathcal{A}_{\mathbb{A}}^{(g)}}{\mathcal{R}^4}
 & =\int \frac{d\ell}{\prod_I(\ell^I)^2}  \int_{{\mathcal M}_{0,4+2g}} \hspace{-10pt} c^{(g)}\big({\mathcal J}^{(g)}\mathcal{Y}^{(g)}\big)^2  \prod_{A=1}^{4+2g}\bar\delta(\mathcal{E}_A)\   \,.
\end{align}
Here, ${\mathcal M}_{0,4+2g}$ is the moduli space of the Riemann sphere with $4+2g$ marked points, corresponding to 4 external particles and $2g$ `loop marked points', one pair per node as in FIG.~\ref{fig:degen}.
 The factors $c^{(g)}$ and ${\mathcal J}^{(g)}$ arise from the degeneration of ${\mathcal M}_{g,4}$ to ${\mathcal M}_{0,4+2g}$ \cite{Geyer:2018xwu}. We will give an example momentarily. The object $\mathcal{Y}^{(g)}$ in this expression is the limit of $\mathcal{Y}^{(g)}_{\mathbb{A}}$ in the maximal non-separating degeneration. Finally, the delta functions impose the loop-level scattering equations on the nodal sphere, $\mathcal{E}_A=0$, on whose finite set of solutions the moduli-space integral fully localises; in fact, this integral can be understood as a multi-dimensional residue integral. The standard $\mathrm{SL}(2,\mathbb{C})$ reduction of the measure and scattering-equation delta functions is understood.

Let us be more concrete. The degeneration to the $g$-nodal sphere is achieved in a limit involving the diagonal components of the period matrix: $q_{II}=e^{i\pi\Omega_{II}} \to 0\,$.
In this limit, the holomorphic Abelian differentials whose periods define the period matrix acquire simple poles at the corresponding node: with $\sigma \in {\mathbb C}{\mathbb P}^1$,
 \begin{equation}
\omega_I=\frac{\omega_{I^+I^-}}{2\pi i}\,, \quad 
\omega_{I^+I^-}(\sigma)= \frac{(\sigma_{I^+}-\sigma_{I^-})\, d\sigma}{(\sigma-\sigma_{I^+})(\sigma-\sigma_{I^-})}\,,
\end{equation}
where the $\sigma_{I^\pm}$ are the marked points for node $I$. Together with the marked points $\sigma_i$ associated to the four external particles, we have the total of $4+2g$ marked points parametrising ${\mathcal M}_{0,4+2g}$ up to $\mathrm{SL}(2,\mathbb{C})$. For $g\geq 2$, the off-diagonal components of the period matrix are expressed in this limit in terms of cross-ratios of the nodal marked points,
 \begin{equation}
 \label{eq:qij}
q_{IJ}=e^{2i\pi\Omega_{IJ}}=\frac{\sigma_{I^+J^+}\sigma_{I^-J^-}}{\sigma_{I^+J^-}\sigma_{I^-J^+}} \,,
\end{equation}
where we denote $\sigma_{AB}=\sigma_A-\sigma_B$\,. This change of integration variables leads to the $({\mathcal J}^{(g)})^2$ appearing in \eqref{eq:assamp}. One ${\mathcal J}^{(g)}$ arises from the moduli-space measure,
\begin{equation}
\label{eq:defJ}
 \prod_{I<J}\frac{dq_{IJ}}{q_{IJ}} 
 = \frac{{\mathcal J}^{(g)}}{\mathrm{vol\; SL}(2,\mathbb{C})}
 \,, \quad {\mathcal J}^{(g)}={ J}^{(g)} \prod_{I^\pm}d\sigma_{I^\pm}\,,
\end{equation}
while the other arises from rewriting higher-genus scattering equations as nodal sphere ones. Finally, the scattering equations on the nodal sphere are equivalent to the vanishing of a meromorphic quadratic differential $\mathfrak{P}^{(g)}$ with only simple poles, and can be read off from the residues of this differential at the $4+2g$ marked points,
\begin{equation}
 \mathcal{E}_A =\mathrm{Res}_{\sigma_{\!A}}\mathfrak{P}^{(g)} \,.
\end{equation}

The ingredients of \eqref{eq:assamp} can be illustrated with the two-loop example. We have $\,c^{(2)} = 1/(1-q_{12})\,$ \footnote{$c^{(2)}$ is associated to the genus-2 fundamental domain constraint $|q_{12}|<1$. In particular, $c^{(2)}$ arises from relaxing this constraint when degenerating to the sphere \cite{Geyer:2018xwu}.} and
\begin{equation}
\mathfrak{P}^{(2)}=P^2 - \sum_I (\ell^I)^2 \,\omega_{I^+I^-}^2 +(\ell_1^2+\ell_2^2)\,\omega_{1^+1^-}\omega_{2^+2^-}\,,
\end{equation}
where
\begin{equation}
 P_\mu(\sigma) = \ell^I_\mu \,\omega_{I^+I^-}(\sigma) + \sum_i \frac{k_{i\mu}}{\sigma-\sigma_i} \,d\sigma \,.
\end{equation}
Effectively, $\mathfrak{P}^{(g)} $ encodes all the potential loop-integrand propagators in an expression like \eqref{eq:assamp}, while $c^{(g)}$ projects out certain unphysical propagators. These details are not important for this paper, where we are concerned with ${\mathcal J}^{(g)}$ and especially $\mathcal{Y}^{(g)}$. At two loops, we have 
\begin{equation}
J^{(2)}=\frac{1}{\sigma_{1^+2^+}\sigma_{1^+2^-}\sigma_{1^-2^+}\sigma_{1^-2^-}}
\end{equation}
and
\begin{equation}\label{eq:Y_g=1,2}
\mathcal{Y}^{(2)} = \frac1{3} \left( (s_{14}-s_{13})\,\Delta^{(2)}_{12}\Delta^{(2)}_{34} +\mathrm{cyc}(234) \right) \,,
\end{equation}
where we used the determinant
\begin{equation}
\label{eq:Deltag}
 \Delta^{(g)}_{i_1\dots i_g} 
 = \varepsilon^{I_1\dots I_g}\,\omega_{I_1}(\sigma_{i_1})\dots \omega_{I_g}(\sigma_{i_g})
\end{equation}
defined for any $g$.
The expression \eqref{eq:Y_g=1,2} is built from the differentials $\omega_{I}$, which naturally lift from the nodal sphere to become the holomorphic Abelian differentials on the genus-$2$ surface. Indeed, the genus-2 expression is also valid as $\mathcal{Y}^{(2)}_{\mathbb{A}}$ in \eqref{eq:asamp} and, crucially for us, as $\mathcal{Y}^{(2)}_{\mathbb{S}}$ in \eqref{eq:ssamp}. The object $ \Delta^{(g)}$ is a modular form of weight $-1$ at any genus, which at genus 2 gives $\mathcal{Y}^{(2)}_{\mathbb{S}}$ the appropriate weight such that the moduli-space integral is well defined. At three loops, the answer is not as simple as \eqref{eq:Y_g=1,2}: $ \Delta^{(3)}$ still arises \cite{Gomez:2013sla}, but additional ingredients are needed, as discussed e.g.~in \cite{Tourkine:2013rda}, and as we will see here.

\section{$\mathcal{Y}^{(g)}_{\mathbb{S}}$ from BCJ numerators}

Let us present and test our strategy. The steps are to:
\begin{enumerate}
 \item[(i)] take a supergravity loop integrand written in a BCJ double-copy representation,
 \item[(ii)] translate that integrand into the ambitwistor string moduli-space integrand localised on the nodal Riemann sphere, i.e.~obtain $\mathcal{Y}^{(g)}$\,,
 \item[(iii)] uplift that formula to a higher-genus modular form conjecturally valid for the superstring, i.e.~obtain $\mathcal{Y}^{(g)}_{\mathbb{S}}$ such that $\mathcal{Y}^{(g)}_{\mathbb{S}}\to \mathcal{Y}^{(g)}$ as $q_{II}\to0$\,.
\end{enumerate}
With our current understanding, step (iii) relies on an educated guess, as we will exemplify.

Starting with step (i), a BCJ representation is one in which the loop integrand is written in terms of trivalent diagrams, whose numerators are the square of analogous numerators in non-planar SYM obeying the BCJ colour-kinematics duality \cite{Bern:2008qj,Bern:2010ue} \footnote{It should be noted that the supergravity loop integrand in \eqref{eq:assamp}, once the moduli-space integral is performed, is not written in terms of Feynman-like propagators, as in the original BCJ representation \cite{Bern:2010ue}. It is instead written in an alternative representation, which was discovered in the ambitwistor string; see \cite{Geyer:2015bja,He:2015yua,Baadsgaard:2015twa,Geyer:2015jch,Cachazo:2015aol,He:2016mzd,Feng:2016nrf,He:2017spx,Geyer:2017ela,Geyer:2019hnn,Edison:2020uzf,Farrow:2020voh} for discussions. The supergravity BCJ numerators relevant in our two- and three-loop problems are valid in both representations.}. See \cite{Bern:2019prr} for a review of this remarkable construction, which was motivated by the KLT relations of string theory \cite{Kawai:1985xq}. Indeed, there is a large body of work relating this construction to aspects of string theory, e.g.~\cite{BjerrumBohr:2009rd,Stieberger:2009hq,Mafra:2011kj,Mafra:2012kh,Ochirov:2013xba,Mafra:2014gja,He:2015wgf,Mafra:2015vca,Tourkine:2016bak,Hohenegger:2017kqy,Ochirov:2017jby,Tourkine:2019ukp,Mizera:2019gea,Casali:2019ihm,Casali:2020knc,Borsten:2021hua,Bridges:2021ebs}. Step (ii) is based on the connection to the scattering equations story, for which we use the following relation based on a differential form with logarithmic singularities \footnote{The tree-level (${\mathcal J}^{(0)}=1$) version of this relation was revealed in string theory in \cite{Mafra:2011nv,Mafra:2011kj} and in the scattering equations (CHY) formalism in \cite{Cachazo:2013iea}; see also \cite{Azevedo:2018dgo,He:2018pol}. Our higher-loop formula is motivated by its verification at two loops in \cite{Geyer:2019hnn}. The higher-multiplicity extension is trivial.}
\begin{equation} 
\label{eq:expKK}
(2\pi i)^4 \, {\mathcal J}^{(g)}\mathcal{Y}^{(g)} = \sum_{\rho\in S_{2+2g}}\frac{N^{(g)}(1^+,\rho,1^-)}{(1^+,\rho,1^-)}
\;\,\prod_{A=1}^{4+2g}d\sigma_A\,,
\end{equation}
where $(ABC\dots D)=\sigma_{AB}\sigma_{BC}\dots\sigma_{DA}$ is a Parke-Taylor denominator. The BCJ numerators $N^{(g)}$, which depend on a particle ordering, are SYM numerators whose square gives the supergravity numerators; this square effectively translates into the square of ${\mathcal J}^{(g)}\mathcal{Y}^{(g)}$ in \eqref{eq:assamp}. Notice, however, that we have extracted the overall factor ${\mathcal{R}^4}$ in \eqref{eq:assamp}, whose `square root' 
is therefore not included in the SYM numerators. The correspondence between the numerators $N^{(g)}$ and trivalent diagrams is best understood in an explicit example, to be discussed below. Before that, let us make two comments. 
 The first is that the two marked points singled out in \eqref{eq:expKK} were chosen to be $\sigma_{1^\pm}$, but the sum is independent of that choice. The second, for the reader familiar with the scattering equations formalism including the developments \cite{Bjerrum-Bohr:2014qwa,Mizera:2019blq,Kalyanapuram:2021xow,Kalyanapuram:2021vjt}, is that equalities like \eqref{eq:expKK} often hold only when the marked points satisfy the scattering equations (e.g.~for CHY Pfaffians). Here, on the other hand, we propose that \eqref{eq:expKK} defines $\mathcal{Y}^{(g)}$ such that it may be uplifted to the superstring, as happens up to two loops.

\begin{figure}[t]
    \includegraphics[width=0.485\textwidth]{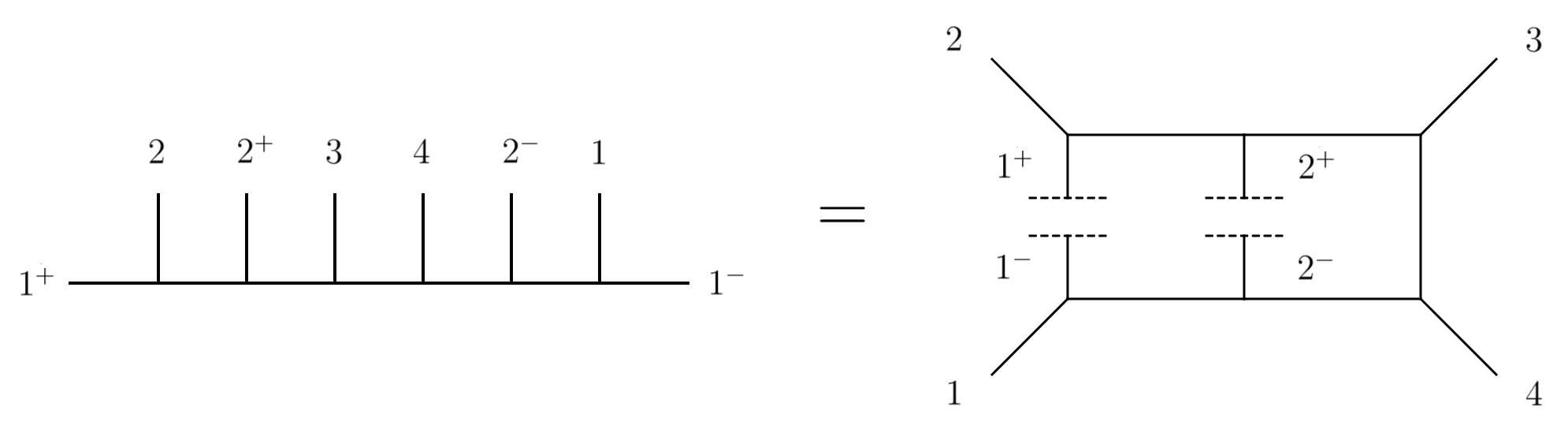}
\caption{Two-loop example. Diagram associated to the numerator $N(1^+,2,2^+,3,4,2^-,1,1^-)$.}
\label{fig:2loopnum}
\end{figure}  

Let us test the strategy at two loops, for which the BCJ representation of the four-point supergravity loop integrand has long been known \cite{Bern:1998ug} \footnote{Though the focus in \cite{Bern:1998ug} was 4D $\mathcal N=8$ supergravity, the four-point loop integrand is valid in $D$ dimensions. We use its 10D specialisation, with the appropriate prefactor ${\mathcal R}^4$. The unregularised 10D supergravity loop {\it integral} is UV divergent.}. The two-loop BCJ numerators can be compactly written as
\begin{equation}
\label{eq:2loopBCJ}
 N^{(2)}(1^+\!,\rho_1,2^{\pm}\!,\rho_2,2^{\mp}\!,\rho_3,1^-)=
    \begin{cases}
     s_{ij} & \rho_2 = \{i,j\}\\
     0 &\text{otherwise}\,.
    \end{cases}
\end{equation}
They correspond to half-ladder diagrams with loop momenta $\pm \ell_1$ at the ends; see FIG.~\ref{fig:2loopnum}.
A standard two-loop diagram is then obtained by gluing the nodal legs, i.e.~$I^+$ with $I^-$. 
Taking the result  \eqref{eq:2loopBCJ} from the literature, it is possible to obtain $\mathcal{Y}^{(2)}$ via \eqref{eq:expKK}. Then, it is both natural and easy to rewrite $\mathcal{Y}^{(2)}$ in the form \eqref{eq:Y_g=1,2}, which as explained earlier can be uplifted to genus 2, matching the superstring result $\mathcal{Y}^{(2)}_{\mathbb{S}}$. This achieves step (iii).

\section{Three Loops}

We now apply our strategy to the much more intricate three-loop case. From the general form of a three-loop field theory integrand, namely the inclusion of the relevant diagram topologies, we can determine $c^{(3)}$ and $\mathfrak{P}^{(3)}$. However, they do not appear in \eqref{eq:expKK}, so they are not important for the goal of this paper \footnote{They will be discussed elsewhere.}.
The important quantities are ${\mathcal J}^{(3)}$ and $\mathcal{Y}^{(3)}$. The Jacobian is straightforwardly obtained from \eqref{eq:defJ} and can be written as
\begin{equation}
J^{(3)} =  J_{\text{hyp}}\, \frac{\prod_I\sigma_{I^+I^-}}{\prod_{I< J}\,\sigma_{I^+J^+}\sigma_{I^-J^-}\sigma_{I^+J^-}\sigma_{I^-J^+}}  \,, 
\end{equation}
where in the factor
\begin{equation}
\label{eq:Jhyp}
  J_{\text{hyp}}  = \sigma_{1^+2^-}\sigma_{2^+3^-}\sigma_{3^+1^-}-\sigma_{1^+3^-}\sigma_{3^+2^-}\sigma_{2^+1^-}
 \end{equation}
the subscript refers to {\it hyperelliptic}, as we will explain.

We can now determine $\mathcal{Y}^{(3)}$ using \eqref{eq:expKK}. The right-hand side is obtained from the known BCJ representation of the three-loop supergravity integrand, a landmark application of the double copy \cite{Bern:2010ue} \footnote{The loop integrand in \cite{Bern:2010ue} was constructed using $D$-dimensional unitarity cuts. We use its 10D specialisation, with the appropriate prefactor ${\mathcal R}^4$.}. The BCJ numerators, listed in table I of \cite{Bern:2010ue}, are not as simple as at two loops and depend linearly on the loop momenta, e.g. \footnote{Our convention for the external momenta is that they are incoming, whereas the convention in \cite{Bern:2010ue} was that they are outgoing. This affects the sign of the term linear in the loop momenta.}
\begin{align}
& N(1^+,1,2,2^+,3,3^+,2^-,4,3^-,1^-) = \frac1{3}\,s_{12}(s_{12}-s_{14}) \nonumber \\ 
&+ \frac{2}{3}\,  \ell^{1}\cdot\big(k_2(s_{13}-s_{14}) +k_3(s_{13}-s_{12})+k_4(s_{12}-s_{14}) \big)
\,. \nonumber
\end{align}
Via \eqref{eq:expKK}, this property implies
\begin{equation}
\label{eq:Y3}
2\pi i\, \mathcal{Y}^{(3)}_{\mathbb{S}}= \mathcal{Y}_0 +  2\pi i\, \ell^I_\mu  \, \mathcal{Y}_I^\mu \,,
\end{equation}
where the factors were chosen for later convenience.
We write our results already in uplifted form, i.e.~for $\mathcal{Y}^{(3)}_{\mathbb{S}}$ (which we claim is $\mathcal{Y}^{(3)}_{\mathbb{A}}$) instead of its degeneration $\mathcal{Y}^{(3)}$. To determine $\mathcal{Y}^{(3)}_{\mathbb{S}}$, we construct a well-motivated ansatz with the required modular weight of $-1$, and fix the coefficients of that ansatz by matching numerically the degeneration limit to \eqref{eq:expKK}. This requires expanding in the degeneration parameters the Jacobi theta functions which define various objects, a straightforward if computationally heavy procedure. 

The second term in \eqref{eq:Y3} is the easiest: we can write
\begin{equation}
\label{eq:Yloop}
\mathcal{Y}^\mu_I = \frac{2}{3} \left( \alpha_1^\mu \, \omega_I(z_1)\Delta^{(3)}_{234} +\mathrm{cyc}(1234) \right) \,,
\end{equation}
with\,  $\alpha_1^\mu = k_2^\mu\,\left(k_3-k_4\right)\cdot k_1+\mathrm{cyc}(234)$ . All the ingredients have been introduced previously.

The object $\mathcal{Y}_0$ is more involved. It is convenient to extricate the kinematic dependence by writing
\begin{equation}
\mathcal{Y}_0 = s_{13}s_{14}\, Y_{12,34} + \mathrm{cyc}(234)\,,
\end{equation}
where $Y_{12,34}$ is independent of the $s_{ij}$ and is symmetric when exchanging: $z_1\leftrightarrow z_2$, $z_3\leftrightarrow z_4$, $\{z_1,z_2\}\leftrightarrow\{z_3,z_4\}$. Let us first state the result and then discuss it \footnote{The factor $(2\pi i)^{-3}$ in the second term of \eqref{eq:Y1234}, introduced in the arXiv version v3, corrects a typo in earlier versions, including the version published in PRL.}: 
\begin{equation}
\label{eq:Y1234}
Y_{12,34} = \frac1{3}\, {\mathcal D}_{12,34} - \frac1{15\,(2\pi i)^3\,\Psi_9} \left( {\mathcal S}_{12,34}^{(a)} -\frac1{8}\,{\mathcal S}_{12,34}^{(b)}\right)
\,,
\end{equation}
where
\begin{align}
\label{eq:wDelta}
{\mathcal D}_{12,34} &= \omega_{3,4}(z_1)\Delta^{(3)}_{234} + \omega_{3,4}(z_2)\Delta^{(3)}_{134} \nonumber \\
&\quad + \omega_{1,2}(z_3)\Delta^{(3)}_{412} + \omega_{1,2}(z_4)\Delta^{(3)}_{312} \,, \\
{} \nonumber\\
\label{eq:Szegoa}
{\mathcal S}_{12,34}^{(a)} &=  \! \sum_\delta  \Xi_8[\delta] \Big( S_\delta(z_1,z_2)S_\delta(z_2,z_3)S_\delta(z_3,z_4)S_\delta(z_4,z_1) \nonumber \\
&  \!\!\!\!\!\! + S_\delta(z_2,z_1)S_\delta(z_1,z_3)S_\delta(z_3,z_4)S_\delta(z_4,z_2) \Big) \,,\\
{} \nonumber\\
\label{eq:Szegob}
{\mathcal S}_{12,34}^{(b)} &=  \sum_\delta \Xi_8[\delta]  \, S_\delta(z_1,z_2)^2 S_\delta(z_3,z_4)^2  \,.
\end{align}

Starting with the expression \eqref{eq:wDelta}, the object $\omega_{i,j}(z_k)$ is the normalised Abelian differential of the third kind, whose degeneration limit is
 \begin{equation}
\omega_{i,j}(\sigma)= \frac{\sigma_{ij}}{(\sigma-\sigma_{i})(\sigma-\sigma_{j})}\, d\sigma \,.
\end{equation}
A consistency check is that the contribution \eqref{eq:wDelta}, including the kinematic coefficient, is completely fixed by \eqref{eq:Yloop}. This follows from the condition of `homology invariance': distinct choices of homology cycles of the Riemann surface with respect to the marked points $z_i$ obey monodromy relations dictated by the chiral splitting procedure \cite{DHoker:1989cxq}, and this connects the two contributions \footnote{In summary, if a marked point $z_i$ is shifted by a `B-cycle', (i) the loop momentum associated to that cycle is shifted by $-k_i$ and (ii) the Abelian differential of the third kind has non-trivial monodromy. These two effects combine precisely to achieve homology invariance. See the very clear discussion for the two-loop five-point amplitude in \cite{DHoker:2020prr}, where the objects $g^I_{i,j}$ relate to $\omega_{3,4}(z_1)$ as $\omega_{3,4}(z_1)=(g^I_{1,3}-g^I_{1,4})\omega_I(z_1)$}.

The contributions \eqref{eq:Szegoa} and \eqref{eq:Szegob} are more elaborate, but the structure is familiar from the RNS formalism \cite{DHoker:1988ta,DHoker:2002hof,DHoker:2001kkt,DHoker:2001qqx,DHoker:2001foj,DHoker:2001jaf,DHoker:2005dys,DHoker:2005vch}. The sums are over the 36 even spin structures at genus 3, labelled by $\delta$, and the objects $S_\delta(z_i,z_j)$ are the Szeg\H{o} kernels arising from the OPEs of worldsheet fermions. The `chiral measure' $\Xi_8[\delta]/\Psi_9$ is the crucial ingredient. Here, $\Psi_9=\sqrt{-\prod_\delta \theta[\delta](0)}$ is a modular form of weight 9 (note our non-standard definition for the sign), defined in terms of the even Jacobi theta functions. The general properties of the chiral measure were described in \cite{DHoker:2004qhf,DHoker:2004fcs} and the precise definition of $\Xi_8[\delta]$ was given in \cite{Cacciatori:2008ay}. It is a sophisticated definition, so we will not repeat it here; we found ref.~\cite{tsuyumine1986} very helpful. The RNS derivation of this measure remains obscure; see appendix C of \cite{Witten:2015hwa}.

In the degeneration limit $q_{II}\to0$, $\Psi_9$ vanishes with leading behaviour $\Psi_9=(\prod_I q_{II}^2)\,\psi_9+\ldots$ ,
\begin{equation}
\psi_9 = 2^{14}\, J_{\text{hyp}}\; \frac{\left(\prod_I\sigma_{I^+I^-}\right)^3}{\prod_{I< J}\,\sigma_{I^+J^+}\sigma_{I^-J^-}\sigma_{I^+J^-}\sigma_{I^-J^+}} \,,
\end{equation}
where $J_{\text{hyp}}$ is given in \eqref{eq:Jhyp}. 
It is opportune to note that only a codimension-1${}_{\mathbb{C}}$ subset of genus-3 Riemann surfaces are hyperelliptic (whereas for $g\leq2$ all surfaces are), and these are precisely identified by the vanishing of $\Psi_9$ \footnote{and the non-vanishing of another modular form called $\Sigma_{140}$ in the classical reference \cite{Igusa}.}. The condition $J_{\text{hyp}}=0$ identifies hyperelliptic surfaces in the degeneration limit. The factors of $J_{\text{hyp}}$ in $J^{(3)}$ and in $1/\Psi_9$ cancel, such that ${\mathcal J}^{(3)}\mathcal{Y}^{(3)}$ does not vanish in the hyperelliptic sector.

The sums \eqref{eq:Szegoa} and \eqref{eq:Szegob}, which are modular forms of weight 8, vanish in the degeneration limit in a manner analogous to $\Psi_9$, so that the ratio appearing in \eqref{eq:Y1234} yields a finite result on the nodal sphere  \footnote{One may also ask why a sum with  
$$
S_\delta(z_1,z_3)^2 S_\delta(z_2,z_4)^2 
+ S_\delta(z_1,z_4)^2 S_\delta(z_2,z_3)^2
$$
is absent from our result, since it has the correct symmetries. It turns out that this sum gives precisely twice the sum \eqref{eq:Szegob}.}. As consistency checks on our implementation of the chiral measure, we verified to order $O(q_{II}^2)$ the following identities (respectively, from \cite{Cacciatori:2008ay,Grushevsky:2008qp,Matone:2008td}):
\begin{align}
& \sum_\delta \Xi_8[\delta]  =0 \,, \qquad  \sum_\delta \Xi_8[\delta]  \, S_\delta(z_1,z_2)^2  =0 \,, \nonumber \\
& \sum_\delta \Xi_8[\delta]  \, S_\delta(z_1,z_2) S_\delta(z_2,z_3) S_\delta(z_3,z_1)= C\, \Psi_9 \, \Delta^{(3)}_{123} \,, \nonumber
\end{align}
where we determined the previously unknown coefficient $C=15\,(2\pi i)^3$. We could not find simplified expressions for \eqref{eq:Szegoa} and \eqref{eq:Szegob}; they are not proportional to $\Psi_9$, i.e.~not proportional to $J_{\text{hyp}}$ in the degeneration limit.

Comparing our result to the pure spinor computation of \cite{Gomez:2013sla}, the latter was restricted to part of the correlator and was not manifestly modular invariant, but appears to be consistent at least with \eqref{eq:Yloop}. The main goal of \cite{Gomez:2013sla}, for which the partial computation was sufficient, was to match a prediction from S-duality \cite{Green:2005ba} for the low-energy amplitude, where the overall normalisation is important. We neglected the normalisation here, and leave this aspect and a proper comparison to \cite{Gomez:2013sla} for future work. Due to manifest supersymmetry, the splitting of spin structures does not arise in the pure spinor approach \cite{Berkovits:2000fe,Berkovits:2001rb,Berkovits:2002zk,Berkovits:2004px,Berkovits:2005df,Berkovits:2005ng}, so this approach may be helpful in simplifying the sums seen above.

\section{Discussion}

We have constructed a conjectured expression for the three-loop four-point amplitude of massless states in the type II superstring. The crucial ingredient is the chiral half-integrand \eqref{eq:Y3}. As at two loops \cite{DHoker:2005vch,DHoker:2020prr}, this object can also in principle be imported into the Heterotic superstring, paired with a bosonic counterpart.

In place of a first-principles worldsheet calculation, we wrote down an ansatz inspired by insights from the RNS and pure spinor formalisms, and then constrained that ansatz using supergravity data mined with modern amplitudes techniques. 
Our focus was on briefly delineating a strategy, with very concrete results. Additional technical details will be presented elsewhere. We hope that our conjecture can guide rigorous derivations using established worldsheet methods. Alternatively, in the spirit of the amplitudes programme, perhaps the proof can follow from a set of basic constraints, such as unitarity.

Natural future directions are: the study of the moduli-space integration in the low-energy limit, building on \cite{DHoker:2005jhf,Gomez:2013sla,DHoker:2013fcx,DHoker:2014oxd,DHoker:2020tcq}, which is newly motivated by beautiful advances in the non-perturbative amplitudes bootstrap \cite{Guerrieri:2021ivu}; and the consideration of higher-point \cite{Tsuchiya:1988va,Richards:2008jg,Mafra:2012kh,Tsuchiya:2012nf,Green:2013bza,Mafra:2014gja,Mafra:2015vca,Mafra:2016nwr,Mafra:2018nla,Mafra:2018pll,Mafra:2018qqe,DHoker:2020prr,DHoker:2020tcq} or higher-loop \cite{DHoker:2004fcs,Cacciatori:2008ay,Grushevsky:2008zm,Cacciatori:2008pj,SalvatiManni:2008qa,Morozov:2008wz,Grushevsky:2008zp,Matone:2010yv,Matone:2005vm} amplitudes. We expect our strategy to prove useful, not least because there are BCJ numerators for ${\mathcal N}=8$ supergravity up to five loops \cite{Bern:2012uf,Bern:2017yxu,Bern:2017ucb}, although the five-loop case required a generalisation of this representation. 
Also at this loop order, the relation between supermoduli space and ordinary moduli space becomes more intricate \cite{Donagi:2013dua}, calling into question the structure of our starting point \eqref{eq:ssamp}. The interplay between field theory and string theory amplitudes continues to present us with many challenges and fruitful surprises.

\vspace{0.3cm}
\noindent
{\textbf{Note}}  As this work was concluded, it came to our knowledge that the authors of \cite{DHoker:2020prr} have independently constructed the contribution to the half-integrand that is linear in the loop momenta, equation \eqref{eq:Yloop}.
       
\vspace{0.3cm}
\begin{acknowledgments}
\noindent
{\textbf{Acknowledgements}} We thank Eric D'Hoker, Carlos Mafra, Boris Pioline, Rodolfo Russo, Oliver Schlotterer and Edward Witten for comments. YG is supported by the CUniverse research promotion project ``Toward World-class Fundamental Physics" of Chulalongkorn University (grant CUAASC). RM and RSM are supported by the Royal Society via a University Research Fellowship and a Studentship Grant, respectively.
\end{acknowledgments}

\bibliography{twistor-bib}

\begin{thebibliography}{122}%
\makeatletter
\providecommand \@ifxundefined [1]{%
 \@ifx{#1\undefined}
}%
\providecommand \@ifnum [1]{%
 \ifnum #1\expandafter \@firstoftwo
 \else \expandafter \@secondoftwo
 \fi
}%
\providecommand \@ifx [1]{%
 \ifx #1\expandafter \@firstoftwo
 \else \expandafter \@secondoftwo
 \fi
}%
\providecommand \natexlab [1]{#1}%
\providecommand \enquote  [1]{``#1''}%
\providecommand \bibnamefont  [1]{#1}%
\providecommand \bibfnamefont [1]{#1}%
\providecommand \citenamefont [1]{#1}%
\providecommand \href@noop [0]{\@secondoftwo}%
\providecommand \href [0]{\begingroup \@sanitize@url \@href}%
\providecommand \@href[1]{\@@startlink{#1}\@@href}%
\providecommand \@@href[1]{\endgroup#1\@@endlink}%
\providecommand \@sanitize@url [0]{\catcode `\\12\catcode `\$12\catcode
  `\&12\catcode `\#12\catcode `\^12\catcode `\_12\catcode `\%12\relax}%
\providecommand \@@startlink[1]{}%
\providecommand \@@endlink[0]{}%
\providecommand \url  [0]{\begingroup\@sanitize@url \@url }%
\providecommand \@url [1]{\endgroup\@href {#1}{\urlprefix }}%
\providecommand \urlprefix  [0]{URL }%
\providecommand \Eprint [0]{\href }%
\providecommand \doibase [0]{https://doi.org/}%
\providecommand \selectlanguage [0]{\@gobble}%
\providecommand \bibinfo  [0]{\@secondoftwo}%
\providecommand \bibfield  [0]{\@secondoftwo}%
\providecommand \translation [1]{[#1]}%
\providecommand \BibitemOpen [0]{}%
\providecommand \bibitemStop [0]{}%
\providecommand \bibitemNoStop [0]{.\EOS\space}%
\providecommand \EOS [0]{\spacefactor3000\relax}%
\providecommand \BibitemShut  [1]{\csname bibitem#1\endcsname}%
\let\auto@bib@innerbib\@empty
\bibitem [{\citenamefont {Veneziano}(1968)}]{Veneziano:1968yb}%
  \BibitemOpen
  \bibfield  {author} {\bibinfo {author} {\bibfnamefont {G.}~\bibnamefont
  {Veneziano}},\ }\bibfield  {title} {\bibinfo {title} {{Construction of a
  crossing - symmetric, Regge behaved amplitude for linearly rising
  trajectories}},\ }\href {https://doi.org/10.1007/BF02824451} {\bibfield
  {journal} {\bibinfo  {journal} {Nuovo Cim. A}\ }\textbf {\bibinfo {volume}
  {57}},\ \bibinfo {pages} {190} (\bibinfo {year} {1968})}\BibitemShut
  {NoStop}%
\bibitem [{\citenamefont {Green}\ and\ \citenamefont
  {Schwarz}(1982)}]{Green:1981yb}%
  \BibitemOpen
  \bibfield  {author} {\bibinfo {author} {\bibfnamefont {M.~B.}\ \bibnamefont
  {Green}}\ and\ \bibinfo {author} {\bibfnamefont {J.~H.}\ \bibnamefont
  {Schwarz}},\ }\bibfield  {title} {\bibinfo {title} {{Supersymmetrical String
  Theories}},\ }\href {https://doi.org/10.1016/0370-2693(82)91110-8} {\bibfield
   {journal} {\bibinfo  {journal} {Phys. Lett. B}\ }\textbf {\bibinfo {volume}
  {109}},\ \bibinfo {pages} {444} (\bibinfo {year} {1982})}\BibitemShut
  {NoStop}%
\bibitem [{\citenamefont {Schwarz}(1982)}]{Schwarz:1982jn}%
  \BibitemOpen
  \bibfield  {author} {\bibinfo {author} {\bibfnamefont {J.~H.}\ \bibnamefont
  {Schwarz}},\ }\bibfield  {title} {\bibinfo {title} {{Superstring Theory}},\
  }\href {https://doi.org/10.1016/0370-1573(82)90087-4} {\bibfield  {journal}
  {\bibinfo  {journal} {Phys. Rept.}\ }\textbf {\bibinfo {volume} {89}},\
  \bibinfo {pages} {223} (\bibinfo {year} {1982})}\BibitemShut {NoStop}%
\bibitem [{\citenamefont {D'Hoker}\ and\ \citenamefont
  {Phong}(2005{\natexlab{a}})}]{DHoker:2005vch}%
  \BibitemOpen
  \bibfield  {author} {\bibinfo {author} {\bibfnamefont {E.}~\bibnamefont
  {D'Hoker}}\ and\ \bibinfo {author} {\bibfnamefont {D.~H.}\ \bibnamefont
  {Phong}},\ }\bibfield  {title} {\bibinfo {title} {{Two-loop superstrings VI:
  Non-renormalization theorems and the 4-point function}},\ }\href
  {https://doi.org/10.1016/j.nuclphysb.2005.02.043} {\bibfield  {journal}
  {\bibinfo  {journal} {Nucl. Phys.}\ }\textbf {\bibinfo {volume} {B715}},\
  \bibinfo {pages} {3} (\bibinfo {year} {2005}{\natexlab{a}})},\ \Eprint
  {https://arxiv.org/abs/hep-th/0501197} {arXiv:hep-th/0501197 [hep-th]}
  \BibitemShut {NoStop}%
\bibitem [{\citenamefont {Berkovits}(2006)}]{Berkovits:2005df}%
  \BibitemOpen
  \bibfield  {author} {\bibinfo {author} {\bibfnamefont {N.}~\bibnamefont
  {Berkovits}},\ }\bibfield  {title} {\bibinfo {title} {{Super-Poincare
  Covariant Two-Loop Superstring Amplitudes}},\ }\href
  {https://doi.org/10.1088/1126-6708/2006/01/005} {\bibfield  {journal}
  {\bibinfo  {journal} {JHEP}\ }\textbf {\bibinfo {volume} {01}},\ \bibinfo
  {pages} {005}},\ \Eprint {https://arxiv.org/abs/hep-th/0503197}
  {arXiv:hep-th/0503197 [hep-th]} \BibitemShut {NoStop}%
\bibitem [{\citenamefont {Berkovits}\ and\ \citenamefont
  {Mafra}(2006)}]{Berkovits:2005ng}%
  \BibitemOpen
  \bibfield  {author} {\bibinfo {author} {\bibfnamefont {N.}~\bibnamefont
  {Berkovits}}\ and\ \bibinfo {author} {\bibfnamefont {C.~R.}\ \bibnamefont
  {Mafra}},\ }\bibfield  {title} {\bibinfo {title} {{Equivalence of Two-Loop
  Superstring Amplitudes in the Pure Spinor and RNS Formalisms}},\ }\href
  {https://doi.org/10.1103/PhysRevLett.96.011602} {\bibfield  {journal}
  {\bibinfo  {journal} {Phys. Rev. Lett.}\ }\textbf {\bibinfo {volume} {96}},\
  \bibinfo {pages} {011602} (\bibinfo {year} {2006})},\ \Eprint
  {https://arxiv.org/abs/hep-th/0509234} {arXiv:hep-th/0509234 [hep-th]}
  \BibitemShut {NoStop}%
\bibitem [{\citenamefont {D'Hoker}\ and\ \citenamefont
  {Phong}(2005{\natexlab{b}})}]{DHoker:2004fcs}%
  \BibitemOpen
  \bibfield  {author} {\bibinfo {author} {\bibfnamefont {E.}~\bibnamefont
  {D'Hoker}}\ and\ \bibinfo {author} {\bibfnamefont {D.~H.}\ \bibnamefont
  {Phong}},\ }\bibfield  {title} {\bibinfo {title} {{Asyzygies, modular forms,
  and the superstring measure II}},\ }\href
  {https://doi.org/10.1016/j.nuclphysb.2004.12.020} {\bibfield  {journal}
  {\bibinfo  {journal} {Nucl. Phys. B}\ }\textbf {\bibinfo {volume} {710}},\
  \bibinfo {pages} {83} (\bibinfo {year} {2005}{\natexlab{b}})},\ \Eprint
  {https://arxiv.org/abs/hep-th/0411182} {arXiv:hep-th/0411182} \BibitemShut
  {NoStop}%
\bibitem [{\citenamefont {D'Hoker}\ and\ \citenamefont
  {Phong}(2005{\natexlab{c}})}]{DHoker:2004qhf}%
  \BibitemOpen
  \bibfield  {author} {\bibinfo {author} {\bibfnamefont {E.}~\bibnamefont
  {D'Hoker}}\ and\ \bibinfo {author} {\bibfnamefont {D.~H.}\ \bibnamefont
  {Phong}},\ }\bibfield  {title} {\bibinfo {title} {{Asyzygies, modular forms,
  and the superstring measure. I.}},\ }\href
  {https://doi.org/10.1016/j.nuclphysb.2004.12.021} {\bibfield  {journal}
  {\bibinfo  {journal} {Nucl. Phys. B}\ }\textbf {\bibinfo {volume} {710}},\
  \bibinfo {pages} {58} (\bibinfo {year} {2005}{\natexlab{c}})},\ \Eprint
  {https://arxiv.org/abs/hep-th/0411159} {arXiv:hep-th/0411159} \BibitemShut
  {NoStop}%
\bibitem [{\citenamefont {Cacciatori}\ \emph
  {et~al.}(2008{\natexlab{a}})\citenamefont {Cacciatori}, \citenamefont
  {Dalla~Piazza},\ and\ \citenamefont {van Geemen}}]{Cacciatori:2008ay}%
  \BibitemOpen
  \bibfield  {author} {\bibinfo {author} {\bibfnamefont {S.~L.}\ \bibnamefont
  {Cacciatori}}, \bibinfo {author} {\bibfnamefont {F.}~\bibnamefont
  {Dalla~Piazza}},\ and\ \bibinfo {author} {\bibfnamefont {B.}~\bibnamefont
  {van Geemen}},\ }\bibfield  {title} {\bibinfo {title} {{Modular Forms and
  Three Loop Superstring Amplitudes}},\ }\href
  {https://doi.org/10.1016/j.nuclphysb.2008.03.007} {\bibfield  {journal}
  {\bibinfo  {journal} {Nucl. Phys. B}\ }\textbf {\bibinfo {volume} {800}},\
  \bibinfo {pages} {565} (\bibinfo {year} {2008}{\natexlab{a}})},\ \Eprint
  {https://arxiv.org/abs/0801.2543} {arXiv:0801.2543 [hep-th]} \BibitemShut
  {NoStop}%
\bibitem [{\citenamefont {Gomez}\ and\ \citenamefont
  {Mafra}(2013)}]{Gomez:2013sla}%
  \BibitemOpen
  \bibfield  {author} {\bibinfo {author} {\bibfnamefont {H.}~\bibnamefont
  {Gomez}}\ and\ \bibinfo {author} {\bibfnamefont {C.~R.}\ \bibnamefont
  {Mafra}},\ }\bibfield  {title} {\bibinfo {title} {{The closed-string 3-loop
  amplitude and S-duality}},\ }\href {https://doi.org/10.1007/JHEP10(2013)217}
  {\bibfield  {journal} {\bibinfo  {journal} {JHEP}\ }\textbf {\bibinfo
  {volume} {10}},\ \bibinfo {pages} {217}},\ \Eprint
  {https://arxiv.org/abs/1308.6567} {arXiv:1308.6567 [hep-th]} \BibitemShut
  {NoStop}%
\bibitem [{\citenamefont {Green}\ \emph {et~al.}(1982)\citenamefont {Green},
  \citenamefont {Schwarz},\ and\ \citenamefont {Brink}}]{Green:1982sw}%
  \BibitemOpen
  \bibfield  {author} {\bibinfo {author} {\bibfnamefont {M.~B.}\ \bibnamefont
  {Green}}, \bibinfo {author} {\bibfnamefont {J.~H.}\ \bibnamefont {Schwarz}},\
  and\ \bibinfo {author} {\bibfnamefont {L.}~\bibnamefont {Brink}},\ }\bibfield
   {title} {\bibinfo {title} {{N=4 Yang-Mills and N=8 Supergravity as Limits of
  String Theories}},\ }\href {https://doi.org/10.1016/0550-3213(82)90336-4}
  {\bibfield  {journal} {\bibinfo  {journal} {Nucl. Phys.}\ }\textbf {\bibinfo
  {volume} {B198}},\ \bibinfo {pages} {474} (\bibinfo {year}
  {1982})}\BibitemShut {NoStop}%
\bibitem [{\citenamefont {Bern}\ \emph {et~al.}(2010)\citenamefont {Bern},
  \citenamefont {Carrasco},\ and\ \citenamefont {Johansson}}]{Bern:2010ue}%
  \BibitemOpen
  \bibfield  {author} {\bibinfo {author} {\bibfnamefont {Z.}~\bibnamefont
  {Bern}}, \bibinfo {author} {\bibfnamefont {J.~J.~M.}\ \bibnamefont
  {Carrasco}},\ and\ \bibinfo {author} {\bibfnamefont {H.}~\bibnamefont
  {Johansson}},\ }\bibfield  {title} {\bibinfo {title} {{Perturbative Quantum
  Gravity as a Double Copy of Gauge Theory}},\ }\href
  {https://doi.org/10.1103/PhysRevLett.105.061602} {\bibfield  {journal}
  {\bibinfo  {journal} {Phys.Rev.Lett.}\ }\textbf {\bibinfo {volume} {105}},\
  \bibinfo {pages} {061602} (\bibinfo {year} {2010})},\ \Eprint
  {https://arxiv.org/abs/1004.0476} {arXiv:1004.0476 [hep-th]} \BibitemShut
  {NoStop}%
\bibitem [{\citenamefont {Green}\ \emph {et~al.}(1988)\citenamefont {Green},
  \citenamefont {Schwarz},\ and\ \citenamefont {Witten}}]{Green:1987sp}%
  \BibitemOpen
  \bibfield  {author} {\bibinfo {author} {\bibfnamefont {M.~B.}\ \bibnamefont
  {Green}}, \bibinfo {author} {\bibfnamefont {J.~H.}\ \bibnamefont {Schwarz}},\
  and\ \bibinfo {author} {\bibfnamefont {E.}~\bibnamefont {Witten}},\ }\href
  {http://www.cambridge.org/us/academic/subjects/physics/theoretical-physics-and-mathematical-physics/superstring-theory-volume-1}
  {\emph {\bibinfo {title} {{SUPERSTRING THEORY. VOL. 1: INTRODUCTION}}}},\
  Cambridge Monographs on Mathematical Physics\ (\bibinfo {year}
  {1988})\BibitemShut {NoStop}%
\bibitem [{Note1()}]{Note1}%
  \BibitemOpen
  \bibinfo {note} {Beyond three loops, the integration over the period matrix
  $\Omega _{IJ}$ must be restricted due to the Schottky problem. Beyond four
  loops, the delicate issue of non-projectedness of supermoduli space is also
  known to arise \cite {Donagi:2013dua}.}\BibitemShut {Stop}%
\bibitem [{\citenamefont {D'Hoker}\ and\ \citenamefont
  {Phong}(1988)}]{DHoker:1988ta}%
  \BibitemOpen
  \bibfield  {author} {\bibinfo {author} {\bibfnamefont {E.}~\bibnamefont
  {D'Hoker}}\ and\ \bibinfo {author} {\bibfnamefont {D.~H.}\ \bibnamefont
  {Phong}},\ }\bibfield  {title} {\bibinfo {title} {{The Geometry of String
  Perturbation Theory}},\ }\href {https://doi.org/10.1103/RevModPhys.60.917}
  {\bibfield  {journal} {\bibinfo  {journal} {Rev. Mod. Phys.}\ }\textbf
  {\bibinfo {volume} {60}},\ \bibinfo {pages} {917} (\bibinfo {year}
  {1988})}\BibitemShut {NoStop}%
\bibitem [{\citenamefont {D'Hoker}\ and\ \citenamefont
  {Phong}(1989)}]{DHoker:1989cxq}%
  \BibitemOpen
  \bibfield  {author} {\bibinfo {author} {\bibfnamefont {E.}~\bibnamefont
  {D'Hoker}}\ and\ \bibinfo {author} {\bibfnamefont {D.~H.}\ \bibnamefont
  {Phong}},\ }\bibfield  {title} {\bibinfo {title} {{Conformal Scalar Fields
  and Chiral Splitting on Superriemann Surfaces}},\ }\href
  {https://doi.org/10.1007/BF01218413} {\bibfield  {journal} {\bibinfo
  {journal} {Commun. Math. Phys.}\ }\textbf {\bibinfo {volume} {125}},\
  \bibinfo {pages} {469} (\bibinfo {year} {1989})}\BibitemShut {NoStop}%
\bibitem [{Note2()}]{Note2}%
  \BibitemOpen
  \bibinfo {note} {This follows from supersymmetry and is clear if we consider
  factorisable NS-NS external states $\varepsilon _i^{\mu \nu }=\epsilon
  _i^{\mu }\protect \tilde \epsilon _i^{\nu }$. The supersymmetric pre-factor
  is ${\protect \mathcal R}^4(\epsilon ,\protect \tilde \epsilon )={\protect
  \mathcal F}^4(\epsilon ){\protect \mathcal F}^4(\protect \tilde \epsilon )$,
  where ${\protect \mathcal F}^4$ is the pre-factor for the open superstring
  and includes products $\epsilon _i\cdot \epsilon _j$. At three loops and four
  points, a 10D Levi-Civita tensor arising from an odd spin structure's zero
  mode may just about be saturated: $\varepsilon _{10}(k_1,k_2,k_3,\epsilon
  _1,\epsilon _2,\epsilon _3,\epsilon _4,\ell _1,\ell _2,\ell _3)$, but it
  would never give rise to any $\epsilon _i\cdot \epsilon _j$. Moreover, the
  contraction of the two Levi-Civita tensors (with $\epsilon _i$ and with
  $\protect \tilde \epsilon _i$) over three indices after loop integration,
  required for a potentially non-vanishing contribution, yields products
  $\epsilon _i\cdot \protect \tilde \epsilon _j$, inconsistent with ${\protect
  \mathcal F}^4(\epsilon ){\protect \mathcal F}^4(\protect \tilde \epsilon )$.
  This discussion is consistent with the results of \cite
  {Gomez:2013sla,Berkovits:2006vc}.}\BibitemShut {Stop}%
\bibitem [{\citenamefont {Cachazo}\ \emph
  {et~al.}(2014{\natexlab{a}})\citenamefont {Cachazo}, \citenamefont {He},\
  and\ \citenamefont {Yuan}}]{Cachazo:2013hca}%
  \BibitemOpen
  \bibfield  {author} {\bibinfo {author} {\bibfnamefont {F.}~\bibnamefont
  {Cachazo}}, \bibinfo {author} {\bibfnamefont {S.}~\bibnamefont {He}},\ and\
  \bibinfo {author} {\bibfnamefont {E.~Y.}\ \bibnamefont {Yuan}},\ }\bibfield
  {title} {\bibinfo {title} {{Scattering of Massless Particles in Arbitrary
  Dimensions}},\ }\href {https://doi.org/10.1103/PhysRevLett.113.171601}
  {\bibfield  {journal} {\bibinfo  {journal} {Phys.Rev.Lett.}\ }\textbf
  {\bibinfo {volume} {113}},\ \bibinfo {pages} {171601} (\bibinfo {year}
  {2014}{\natexlab{a}})},\ \Eprint {https://arxiv.org/abs/1307.2199}
  {arXiv:1307.2199 [hep-th]} \BibitemShut {NoStop}%
\bibitem [{\citenamefont {Cachazo}\ \emph
  {et~al.}(2014{\natexlab{b}})\citenamefont {Cachazo}, \citenamefont {He},\
  and\ \citenamefont {Yuan}}]{Cachazo:2013iea}%
  \BibitemOpen
  \bibfield  {author} {\bibinfo {author} {\bibfnamefont {F.}~\bibnamefont
  {Cachazo}}, \bibinfo {author} {\bibfnamefont {S.}~\bibnamefont {He}},\ and\
  \bibinfo {author} {\bibfnamefont {E.~Y.}\ \bibnamefont {Yuan}},\ }\bibfield
  {title} {\bibinfo {title} {{Scattering of Massless Particles: Scalars, Gluons
  and Gravitons}},\ }\href {https://doi.org/10.1007/JHEP07(2014)033} {\bibfield
   {journal} {\bibinfo  {journal} {JHEP}\ }\textbf {\bibinfo {volume} {1407}},\
  \bibinfo {pages} {033}},\ \Eprint {https://arxiv.org/abs/1309.0885}
  {arXiv:1309.0885 [hep-th]} \BibitemShut {NoStop}%
\bibitem [{\citenamefont {Witten}(2004)}]{Witten:2003nn}%
  \BibitemOpen
  \bibfield  {author} {\bibinfo {author} {\bibfnamefont {E.}~\bibnamefont
  {Witten}},\ }\bibfield  {title} {\bibinfo {title} {{Perturbative gauge theory
  as a string theory in twistor space}},\ }\href
  {https://doi.org/10.1007/s00220-004-1187-3} {\bibfield  {journal} {\bibinfo
  {journal} {Commun.Math.Phys.}\ }\textbf {\bibinfo {volume} {252}},\ \bibinfo
  {pages} {189} (\bibinfo {year} {2004})},\ \Eprint
  {https://arxiv.org/abs/hep-th/0312171} {arXiv:hep-th/0312171 [hep-th]}
  \BibitemShut {NoStop}%
\bibitem [{\citenamefont {Roiban}\ \emph {et~al.}(2004)\citenamefont {Roiban},
  \citenamefont {Spradlin},\ and\ \citenamefont {Volovich}}]{Roiban:2004yf}%
  \BibitemOpen
  \bibfield  {author} {\bibinfo {author} {\bibfnamefont {R.}~\bibnamefont
  {Roiban}}, \bibinfo {author} {\bibfnamefont {M.}~\bibnamefont {Spradlin}},\
  and\ \bibinfo {author} {\bibfnamefont {A.}~\bibnamefont {Volovich}},\
  }\bibfield  {title} {\bibinfo {title} {{On the tree level S matrix of
  Yang-Mills theory}},\ }\href {https://doi.org/10.1103/PhysRevD.70.026009}
  {\bibfield  {journal} {\bibinfo  {journal} {Phys.Rev.}\ }\textbf {\bibinfo
  {volume} {D70}},\ \bibinfo {pages} {026009} (\bibinfo {year} {2004})},\
  \Eprint {https://arxiv.org/abs/hep-th/0403190} {arXiv:hep-th/0403190
  [hep-th]} \BibitemShut {NoStop}%
\bibitem [{\citenamefont {Adamo}\ \emph {et~al.}(2014)\citenamefont {Adamo},
  \citenamefont {Casali},\ and\ \citenamefont {Skinner}}]{Adamo:2013tsa}%
  \BibitemOpen
  \bibfield  {author} {\bibinfo {author} {\bibfnamefont {T.}~\bibnamefont
  {Adamo}}, \bibinfo {author} {\bibfnamefont {E.}~\bibnamefont {Casali}},\ and\
  \bibinfo {author} {\bibfnamefont {D.}~\bibnamefont {Skinner}},\ }\bibfield
  {title} {\bibinfo {title} {{Ambitwistor strings and the scattering equations
  at one loop}},\ }\href {https://doi.org/10.1007/JHEP04(2014)104} {\bibfield
  {journal} {\bibinfo  {journal} {JHEP}\ }\textbf {\bibinfo {volume} {1404}},\
  \bibinfo {pages} {104}},\ \Eprint {https://arxiv.org/abs/1312.3828}
  {arXiv:1312.3828 [hep-th]} \BibitemShut {NoStop}%
\bibitem [{\citenamefont {Casali}\ and\ \citenamefont
  {Tourkine}(2015)}]{Casali:2014hfa}%
  \BibitemOpen
  \bibfield  {author} {\bibinfo {author} {\bibfnamefont {E.}~\bibnamefont
  {Casali}}\ and\ \bibinfo {author} {\bibfnamefont {P.}~\bibnamefont
  {Tourkine}},\ }\bibfield  {title} {\bibinfo {title} {{Infrared behaviour of
  the one-loop scattering equations and supergravity integrands}},\ }\href
  {https://doi.org/10.1007/JHEP04(2015)013} {\bibfield  {journal} {\bibinfo
  {journal} {JHEP}\ }\textbf {\bibinfo {volume} {1504}},\ \bibinfo {pages}
  {013}},\ \Eprint {https://arxiv.org/abs/1412.3787} {arXiv:1412.3787 [hep-th]}
  \BibitemShut {NoStop}%
\bibitem [{\citenamefont {Adamo}\ and\ \citenamefont
  {Casali}(2015)}]{Adamo:2015hoa}%
  \BibitemOpen
  \bibfield  {author} {\bibinfo {author} {\bibfnamefont {T.}~\bibnamefont
  {Adamo}}\ and\ \bibinfo {author} {\bibfnamefont {E.}~\bibnamefont {Casali}},\
  }\bibfield  {title} {\bibinfo {title} {{Scattering equations, supergravity
  integrands, and pure spinors}},\ }\href
  {https://doi.org/10.1007/JHEP05(2015)120} {\bibfield  {journal} {\bibinfo
  {journal} {JHEP}\ }\textbf {\bibinfo {volume} {1505}},\ \bibinfo {pages}
  {120}},\ \Eprint {https://arxiv.org/abs/1502.06826} {arXiv:1502.06826
  [hep-th]} \BibitemShut {NoStop}%
\bibitem [{\citenamefont {Geyer}\ \emph {et~al.}(2015)\citenamefont {Geyer},
  \citenamefont {Mason}, \citenamefont {Monteiro},\ and\ \citenamefont
  {Tourkine}}]{Geyer:2015bja}%
  \BibitemOpen
  \bibfield  {author} {\bibinfo {author} {\bibfnamefont {Y.}~\bibnamefont
  {Geyer}}, \bibinfo {author} {\bibfnamefont {L.}~\bibnamefont {Mason}},
  \bibinfo {author} {\bibfnamefont {R.}~\bibnamefont {Monteiro}},\ and\
  \bibinfo {author} {\bibfnamefont {P.}~\bibnamefont {Tourkine}},\ }\bibfield
  {title} {\bibinfo {title} {{Loop Integrands for Scattering Amplitudes from
  the Riemann Sphere}},\ }\href
  {https://doi.org/10.1103/PhysRevLett.115.121603} {\bibfield  {journal}
  {\bibinfo  {journal} {Phys. Rev. Lett.}\ }\textbf {\bibinfo {volume} {115}},\
  \bibinfo {pages} {121603} (\bibinfo {year} {2015})},\ \Eprint
  {https://arxiv.org/abs/1507.00321} {arXiv:1507.00321 [hep-th]} \BibitemShut
  {NoStop}%
\bibitem [{\citenamefont {Geyer}\ \emph
  {et~al.}(2016{\natexlab{a}})\citenamefont {Geyer}, \citenamefont {Mason},
  \citenamefont {Monteiro},\ and\ \citenamefont {Tourkine}}]{Geyer:2015jch}%
  \BibitemOpen
  \bibfield  {author} {\bibinfo {author} {\bibfnamefont {Y.}~\bibnamefont
  {Geyer}}, \bibinfo {author} {\bibfnamefont {L.}~\bibnamefont {Mason}},
  \bibinfo {author} {\bibfnamefont {R.}~\bibnamefont {Monteiro}},\ and\
  \bibinfo {author} {\bibfnamefont {P.}~\bibnamefont {Tourkine}},\ }\bibfield
  {title} {\bibinfo {title} {{One-loop amplitudes on the Riemann sphere}},\
  }\href {https://doi.org/10.1007/JHEP03(2016)114} {\bibfield  {journal}
  {\bibinfo  {journal} {JHEP}\ }\textbf {\bibinfo {volume} {03}},\ \bibinfo
  {pages} {114}},\ \Eprint {https://arxiv.org/abs/1511.06315} {arXiv:1511.06315
  [hep-th]} \BibitemShut {NoStop}%
\bibitem [{\citenamefont {Geyer}\ \emph
  {et~al.}(2016{\natexlab{b}})\citenamefont {Geyer}, \citenamefont {Mason},
  \citenamefont {Monteiro},\ and\ \citenamefont {Tourkine}}]{Geyer:2016wjx}%
  \BibitemOpen
  \bibfield  {author} {\bibinfo {author} {\bibfnamefont {Y.}~\bibnamefont
  {Geyer}}, \bibinfo {author} {\bibfnamefont {L.}~\bibnamefont {Mason}},
  \bibinfo {author} {\bibfnamefont {R.}~\bibnamefont {Monteiro}},\ and\
  \bibinfo {author} {\bibfnamefont {P.}~\bibnamefont {Tourkine}},\ }\bibfield
  {title} {\bibinfo {title} {{Two-Loop Scattering Amplitudes from the Riemann
  Sphere}},\ }\href {https://doi.org/10.1103/PhysRevD.94.125029} {\bibfield
  {journal} {\bibinfo  {journal} {Phys. Rev.}\ }\textbf {\bibinfo {volume}
  {D94}},\ \bibinfo {pages} {125029} (\bibinfo {year} {2016}{\natexlab{b}})},\
  \Eprint {https://arxiv.org/abs/1607.08887} {arXiv:1607.08887 [hep-th]}
  \BibitemShut {NoStop}%
\bibitem [{\citenamefont {Geyer}\ and\ \citenamefont
  {Monteiro}(2018{\natexlab{a}})}]{Geyer:2018xwu}%
  \BibitemOpen
  \bibfield  {author} {\bibinfo {author} {\bibfnamefont {Y.}~\bibnamefont
  {Geyer}}\ and\ \bibinfo {author} {\bibfnamefont {R.}~\bibnamefont
  {Monteiro}},\ }\bibfield  {title} {\bibinfo {title} {{Two-Loop Scattering
  Amplitudes from Ambitwistor Strings: from Genus Two to the Nodal Riemann
  Sphere}},\ }\href {https://doi.org/10.1007/JHEP11(2018)008} {\bibfield
  {journal} {\bibinfo  {journal} {JHEP}\ }\textbf {\bibinfo {volume} {11}},\
  \bibinfo {pages} {008}},\ \Eprint {https://arxiv.org/abs/1805.05344}
  {arXiv:1805.05344 [hep-th]} \BibitemShut {NoStop}%
\bibitem [{\citenamefont {Mason}\ and\ \citenamefont
  {Skinner}(2014)}]{Mason:2013sva}%
  \BibitemOpen
  \bibfield  {author} {\bibinfo {author} {\bibfnamefont {L.}~\bibnamefont
  {Mason}}\ and\ \bibinfo {author} {\bibfnamefont {D.}~\bibnamefont
  {Skinner}},\ }\bibfield  {title} {\bibinfo {title} {{Ambitwistor strings and
  the scattering equations}},\ }\href {https://doi.org/10.1007/JHEP07(2014)048}
  {\bibfield  {journal} {\bibinfo  {journal} {JHEP}\ }\textbf {\bibinfo
  {volume} {1407}},\ \bibinfo {pages} {048}},\ \Eprint
  {https://arxiv.org/abs/1311.2564} {arXiv:1311.2564 [hep-th]} \BibitemShut
  {NoStop}%
\bibitem [{Note3()}]{Note3}%
  \BibitemOpen
  \bibinfo {note} {For the delta functions, we use the definition $\protect
  \bar \delta (z)=\protect \bar \partial (1/2\pi i z)$, standard in this
  context}\BibitemShut {NoStop}%
\bibitem [{Note4()}]{Note4}%
  \BibitemOpen
  \bibinfo {note} {$c^{(2)}$ is associated to the genus-2 fundamental domain
  constraint $|q_{12}|<1$. In particular, $c^{(2)}$ arises from relaxing this
  constraint when degenerating to the sphere \cite
  {Geyer:2018xwu}.}\BibitemShut {Stop}%
\bibitem [{\citenamefont {Tourkine}(2017)}]{Tourkine:2013rda}%
  \BibitemOpen
  \bibfield  {author} {\bibinfo {author} {\bibfnamefont {P.}~\bibnamefont
  {Tourkine}},\ }\bibfield  {title} {\bibinfo {title} {{Tropical Amplitudes}},\
  }\href {https://doi.org/10.1007/s00023-017-0560-7} {\bibfield  {journal}
  {\bibinfo  {journal} {Annales Henri Poincare}\ }\textbf {\bibinfo {volume}
  {18}},\ \bibinfo {pages} {2199} (\bibinfo {year} {2017})},\ \Eprint
  {https://arxiv.org/abs/1309.3551} {arXiv:1309.3551 [hep-th]} \BibitemShut
  {NoStop}%
\bibitem [{\citenamefont {Bern}\ \emph {et~al.}(2008)\citenamefont {Bern},
  \citenamefont {Carrasco},\ and\ \citenamefont {Johansson}}]{Bern:2008qj}%
  \BibitemOpen
  \bibfield  {author} {\bibinfo {author} {\bibfnamefont {Z.}~\bibnamefont
  {Bern}}, \bibinfo {author} {\bibfnamefont {J.}~\bibnamefont {Carrasco}},\
  and\ \bibinfo {author} {\bibfnamefont {H.}~\bibnamefont {Johansson}},\
  }\bibfield  {title} {\bibinfo {title} {{New Relations for Gauge-Theory
  Amplitudes}},\ }\href {https://doi.org/10.1103/PhysRevD.78.085011} {\bibfield
   {journal} {\bibinfo  {journal} {Phys.Rev.}\ }\textbf {\bibinfo {volume}
  {D78}},\ \bibinfo {pages} {085011} (\bibinfo {year} {2008})},\ \Eprint
  {https://arxiv.org/abs/0805.3993} {arXiv:0805.3993 [hep-ph]} \BibitemShut
  {NoStop}%
\bibitem [{Note5()}]{Note5}%
  \BibitemOpen
  \bibinfo {note} {It should be noted that the supergravity loop integrand in
  \protect \eqref {eq:assamp}, once the moduli-space integral is performed, is
  not written in terms of Feynman-like propagators, as in the original BCJ
  representation \cite {Bern:2010ue}. It is instead written in an alternative
  representation, which was discovered in the ambitwistor string; see \cite
  {Geyer:2015bja,He:2015yua,Baadsgaard:2015twa,Geyer:2015jch,Cachazo:2015aol,He:2016mzd,Feng:2016nrf,He:2017spx,Geyer:2017ela,Geyer:2019hnn,Edison:2020uzf,Farrow:2020voh}
  for discussions. The supergravity BCJ numerators relevant in our two- and
  three-loop problems are valid in both representations.}\BibitemShut {Stop}%
\bibitem [{\citenamefont {Bern}\ \emph {et~al.}(2019)\citenamefont {Bern},
  \citenamefont {Carrasco}, \citenamefont {Chiodaroli}, \citenamefont
  {Johansson},\ and\ \citenamefont {Roiban}}]{Bern:2019prr}%
  \BibitemOpen
  \bibfield  {author} {\bibinfo {author} {\bibfnamefont {Z.}~\bibnamefont
  {Bern}}, \bibinfo {author} {\bibfnamefont {J.~J.}\ \bibnamefont {Carrasco}},
  \bibinfo {author} {\bibfnamefont {M.}~\bibnamefont {Chiodaroli}}, \bibinfo
  {author} {\bibfnamefont {H.}~\bibnamefont {Johansson}},\ and\ \bibinfo
  {author} {\bibfnamefont {R.}~\bibnamefont {Roiban}},\ }\bibfield  {title}
  {\bibinfo {title} {{The Duality Between Color and Kinematics and its
  Applications}},\ }\href@noop {} {\  (\bibinfo {year} {2019})},\ \Eprint
  {https://arxiv.org/abs/1909.01358} {arXiv:1909.01358 [hep-th]} \BibitemShut
  {NoStop}%
\bibitem [{\citenamefont {Kawai}\ \emph {et~al.}(1986)\citenamefont {Kawai},
  \citenamefont {Lewellen},\ and\ \citenamefont {Tye}}]{Kawai:1985xq}%
  \BibitemOpen
  \bibfield  {author} {\bibinfo {author} {\bibfnamefont {H.}~\bibnamefont
  {Kawai}}, \bibinfo {author} {\bibfnamefont {D.~C.}\ \bibnamefont
  {Lewellen}},\ and\ \bibinfo {author} {\bibfnamefont {S.~H.~H.}\ \bibnamefont
  {Tye}},\ }\bibfield  {title} {\bibinfo {title} {{A Relation Between Tree
  Amplitudes of Closed and Open Strings}},\ }\href
  {https://doi.org/10.1016/0550-3213(86)90362-7} {\bibfield  {journal}
  {\bibinfo  {journal} {Nucl. Phys.}\ }\textbf {\bibinfo {volume} {B269}},\
  \bibinfo {pages} {1} (\bibinfo {year} {1986})}\BibitemShut {NoStop}%
\bibitem [{\citenamefont {Bjerrum-Bohr}\ \emph {et~al.}(2009)\citenamefont
  {Bjerrum-Bohr}, \citenamefont {Damgaard},\ and\ \citenamefont
  {Vanhove}}]{BjerrumBohr:2009rd}%
  \BibitemOpen
  \bibfield  {author} {\bibinfo {author} {\bibfnamefont {N.~E.~J.}\
  \bibnamefont {Bjerrum-Bohr}}, \bibinfo {author} {\bibfnamefont {P.~H.}\
  \bibnamefont {Damgaard}},\ and\ \bibinfo {author} {\bibfnamefont
  {P.}~\bibnamefont {Vanhove}},\ }\bibfield  {title} {\bibinfo {title}
  {{Minimal Basis for Gauge Theory Amplitudes}},\ }\href
  {https://doi.org/10.1103/PhysRevLett.103.161602} {\bibfield  {journal}
  {\bibinfo  {journal} {Phys. Rev. Lett.}\ }\textbf {\bibinfo {volume} {103}},\
  \bibinfo {pages} {161602} (\bibinfo {year} {2009})},\ \Eprint
  {https://arxiv.org/abs/0907.1425} {arXiv:0907.1425 [hep-th]} \BibitemShut
  {NoStop}%
\bibitem [{\citenamefont {Stieberger}(2009)}]{Stieberger:2009hq}%
  \BibitemOpen
  \bibfield  {author} {\bibinfo {author} {\bibfnamefont {S.}~\bibnamefont
  {Stieberger}},\ }\bibfield  {title} {\bibinfo {title} {{Open \& Closed vs.
  Pure Open String Disk Amplitudes}},\ }\href@noop {} {\  (\bibinfo {year}
  {2009})},\ \Eprint {https://arxiv.org/abs/0907.2211} {arXiv:0907.2211
  [hep-th]} \BibitemShut {NoStop}%
\bibitem [{\citenamefont {Mafra}\ \emph {et~al.}(2011)\citenamefont {Mafra},
  \citenamefont {Schlotterer},\ and\ \citenamefont
  {Stieberger}}]{Mafra:2011kj}%
  \BibitemOpen
  \bibfield  {author} {\bibinfo {author} {\bibfnamefont {C.~R.}\ \bibnamefont
  {Mafra}}, \bibinfo {author} {\bibfnamefont {O.}~\bibnamefont {Schlotterer}},\
  and\ \bibinfo {author} {\bibfnamefont {S.}~\bibnamefont {Stieberger}},\
  }\bibfield  {title} {\bibinfo {title} {{Explicit BCJ Numerators from Pure
  Spinors}},\ }\href {https://doi.org/10.1007/JHEP07(2011)092} {\bibfield
  {journal} {\bibinfo  {journal} {JHEP}\ }\textbf {\bibinfo {volume} {07}},\
  \bibinfo {pages} {092}},\ \Eprint {https://arxiv.org/abs/1104.5224}
  {arXiv:1104.5224 [hep-th]} \BibitemShut {NoStop}%
\bibitem [{\citenamefont {Mafra}\ and\ \citenamefont
  {Schlotterer}(2014)}]{Mafra:2012kh}%
  \BibitemOpen
  \bibfield  {author} {\bibinfo {author} {\bibfnamefont {C.~R.}\ \bibnamefont
  {Mafra}}\ and\ \bibinfo {author} {\bibfnamefont {O.}~\bibnamefont
  {Schlotterer}},\ }\bibfield  {title} {\bibinfo {title} {{The Structure of
  n-Point One-Loop Open Superstring Amplitudes}},\ }\href
  {https://doi.org/10.1007/JHEP08(2014)099} {\bibfield  {journal} {\bibinfo
  {journal} {JHEP}\ }\textbf {\bibinfo {volume} {08}},\ \bibinfo {pages}
  {099}},\ \Eprint {https://arxiv.org/abs/1203.6215} {arXiv:1203.6215 [hep-th]}
  \BibitemShut {NoStop}%
\bibitem [{\citenamefont {Ochirov}\ and\ \citenamefont
  {Tourkine}(2014)}]{Ochirov:2013xba}%
  \BibitemOpen
  \bibfield  {author} {\bibinfo {author} {\bibfnamefont {A.}~\bibnamefont
  {Ochirov}}\ and\ \bibinfo {author} {\bibfnamefont {P.}~\bibnamefont
  {Tourkine}},\ }\bibfield  {title} {\bibinfo {title} {{BCJ duality and double
  copy in the closed string sector}},\ }\href
  {https://doi.org/10.1007/JHEP05(2014)136} {\bibfield  {journal} {\bibinfo
  {journal} {JHEP}\ }\textbf {\bibinfo {volume} {05}},\ \bibinfo {pages}
  {136}},\ \Eprint {https://arxiv.org/abs/1312.1326} {arXiv:1312.1326 [hep-th]}
  \BibitemShut {NoStop}%
\bibitem [{\citenamefont {Mafra}\ and\ \citenamefont
  {Schlotterer}(2015{\natexlab{a}})}]{Mafra:2014gja}%
  \BibitemOpen
  \bibfield  {author} {\bibinfo {author} {\bibfnamefont {C.~R.}\ \bibnamefont
  {Mafra}}\ and\ \bibinfo {author} {\bibfnamefont {O.}~\bibnamefont
  {Schlotterer}},\ }\bibfield  {title} {\bibinfo {title} {{Towards one-loop SYM
  amplitudes from the pure spinor BRST cohomology}},\ }\href
  {https://doi.org/10.1002/prop.201400076} {\bibfield  {journal} {\bibinfo
  {journal} {Fortsch. Phys.}\ }\textbf {\bibinfo {volume} {63}},\ \bibinfo
  {pages} {105} (\bibinfo {year} {2015}{\natexlab{a}})},\ \Eprint
  {https://arxiv.org/abs/1410.0668} {arXiv:1410.0668 [hep-th]} \BibitemShut
  {NoStop}%
\bibitem [{\citenamefont {He}\ \emph {et~al.}(2016)\citenamefont {He},
  \citenamefont {Monteiro},\ and\ \citenamefont {Schlotterer}}]{He:2015wgf}%
  \BibitemOpen
  \bibfield  {author} {\bibinfo {author} {\bibfnamefont {S.}~\bibnamefont
  {He}}, \bibinfo {author} {\bibfnamefont {R.}~\bibnamefont {Monteiro}},\ and\
  \bibinfo {author} {\bibfnamefont {O.}~\bibnamefont {Schlotterer}},\
  }\bibfield  {title} {\bibinfo {title} {{String-inspired BCJ numerators for
  one-loop MHV amplitudes}},\ }\href {https://doi.org/10.1007/JHEP01(2016)171}
  {\bibfield  {journal} {\bibinfo  {journal} {JHEP}\ }\textbf {\bibinfo
  {volume} {01}},\ \bibinfo {pages} {171}},\ \Eprint
  {https://arxiv.org/abs/1507.06288} {arXiv:1507.06288 [hep-th]} \BibitemShut
  {NoStop}%
\bibitem [{\citenamefont {Mafra}\ and\ \citenamefont
  {Schlotterer}(2015{\natexlab{b}})}]{Mafra:2015vca}%
  \BibitemOpen
  \bibfield  {author} {\bibinfo {author} {\bibfnamefont {C.~R.}\ \bibnamefont
  {Mafra}}\ and\ \bibinfo {author} {\bibfnamefont {O.}~\bibnamefont
  {Schlotterer}},\ }\bibfield  {title} {\bibinfo {title} {{Berends-Giele
  recursions and the BCJ duality in superspace and components}},\ }\href@noop
  {} {\  (\bibinfo {year} {2015}{\natexlab{b}})},\ \Eprint
  {https://arxiv.org/abs/1510.08846} {arXiv:1510.08846 [hep-th]} \BibitemShut
  {NoStop}%
\bibitem [{\citenamefont {Tourkine}\ and\ \citenamefont
  {Vanhove}(2016)}]{Tourkine:2016bak}%
  \BibitemOpen
  \bibfield  {author} {\bibinfo {author} {\bibfnamefont {P.}~\bibnamefont
  {Tourkine}}\ and\ \bibinfo {author} {\bibfnamefont {P.}~\bibnamefont
  {Vanhove}},\ }\bibfield  {title} {\bibinfo {title} {{Higher-loop amplitude
  monodromy relations in string and gauge theory}},\ }\href
  {https://doi.org/10.1103/PhysRevLett.117.211601} {\bibfield  {journal}
  {\bibinfo  {journal} {Phys. Rev. Lett.}\ }\textbf {\bibinfo {volume} {117}},\
  \bibinfo {pages} {211601} (\bibinfo {year} {2016})},\ \Eprint
  {https://arxiv.org/abs/1608.01665} {arXiv:1608.01665 [hep-th]} \BibitemShut
  {NoStop}%
\bibitem [{\citenamefont {Hohenegger}\ and\ \citenamefont
  {Stieberger}(2017)}]{Hohenegger:2017kqy}%
  \BibitemOpen
  \bibfield  {author} {\bibinfo {author} {\bibfnamefont {S.}~\bibnamefont
  {Hohenegger}}\ and\ \bibinfo {author} {\bibfnamefont {S.}~\bibnamefont
  {Stieberger}},\ }\bibfield  {title} {\bibinfo {title} {{Monodromy Relations
  in Higher-Loop String Amplitudes}},\ }\href
  {https://doi.org/10.1016/j.nuclphysb.2017.09.020} {\bibfield  {journal}
  {\bibinfo  {journal} {Nucl. Phys.}\ }\textbf {\bibinfo {volume} {B925}},\
  \bibinfo {pages} {63} (\bibinfo {year} {2017})},\ \Eprint
  {https://arxiv.org/abs/1702.04963} {arXiv:1702.04963 [hep-th]} \BibitemShut
  {NoStop}%
\bibitem [{\citenamefont {Ochirov}\ \emph {et~al.}(2017)\citenamefont
  {Ochirov}, \citenamefont {Tourkine},\ and\ \citenamefont
  {Vanhove}}]{Ochirov:2017jby}%
  \BibitemOpen
  \bibfield  {author} {\bibinfo {author} {\bibfnamefont {A.}~\bibnamefont
  {Ochirov}}, \bibinfo {author} {\bibfnamefont {P.}~\bibnamefont {Tourkine}},\
  and\ \bibinfo {author} {\bibfnamefont {P.}~\bibnamefont {Vanhove}},\
  }\bibfield  {title} {\bibinfo {title} {{One-loop monodromy relations on
  single cuts}},\ }\href {https://doi.org/10.1007/JHEP10(2017)105} {\bibfield
  {journal} {\bibinfo  {journal} {JHEP}\ }\textbf {\bibinfo {volume} {10}},\
  \bibinfo {pages} {105}},\ \Eprint {https://arxiv.org/abs/1707.05775}
  {arXiv:1707.05775 [hep-th]} \BibitemShut {NoStop}%
\bibitem [{\citenamefont {Tourkine}(2020)}]{Tourkine:2019ukp}%
  \BibitemOpen
  \bibfield  {author} {\bibinfo {author} {\bibfnamefont {P.}~\bibnamefont
  {Tourkine}},\ }\bibfield  {title} {\bibinfo {title} {{Integrands and loop
  momentum in string and field theory}},\ }\href
  {https://doi.org/10.1103/PhysRevD.102.026006} {\bibfield  {journal} {\bibinfo
   {journal} {Phys. Rev. D}\ }\textbf {\bibinfo {volume} {102}},\ \bibinfo
  {pages} {026006} (\bibinfo {year} {2020})},\ \Eprint
  {https://arxiv.org/abs/1901.02432} {arXiv:1901.02432 [hep-th]} \BibitemShut
  {NoStop}%
\bibitem [{\citenamefont {Mizera}(2020{\natexlab{a}})}]{Mizera:2019gea}%
  \BibitemOpen
  \bibfield  {author} {\bibinfo {author} {\bibfnamefont {S.}~\bibnamefont
  {Mizera}},\ }\emph {\bibinfo {title} {{Aspects of Scattering Amplitudes and
  Moduli Space Localization}}},\ \href
  {https://doi.org/10.1007/978-3-030-53010-5} {Ph.D. thesis},\ \bibinfo
  {school} {Princeton, Inst. Advanced Study} (\bibinfo {year}
  {2020}{\natexlab{a}}),\ \Eprint {https://arxiv.org/abs/1906.02099}
  {arXiv:1906.02099 [hep-th]} \BibitemShut {NoStop}%
\bibitem [{\citenamefont {Casali}\ \emph {et~al.}(2019)\citenamefont {Casali},
  \citenamefont {Mizera},\ and\ \citenamefont {Tourkine}}]{Casali:2019ihm}%
  \BibitemOpen
  \bibfield  {author} {\bibinfo {author} {\bibfnamefont {E.}~\bibnamefont
  {Casali}}, \bibinfo {author} {\bibfnamefont {S.}~\bibnamefont {Mizera}},\
  and\ \bibinfo {author} {\bibfnamefont {P.}~\bibnamefont {Tourkine}},\
  }\bibfield  {title} {\bibinfo {title} {{Monodromy relations from twisted
  homology}},\ }\href {https://doi.org/10.1007/JHEP12(2019)087} {\bibfield
  {journal} {\bibinfo  {journal} {JHEP}\ }\textbf {\bibinfo {volume} {12}},\
  \bibinfo {pages} {087}},\ \Eprint {https://arxiv.org/abs/1910.08514}
  {arXiv:1910.08514 [hep-th]} \BibitemShut {NoStop}%
\bibitem [{\citenamefont {Casali}\ \emph {et~al.}(2021)\citenamefont {Casali},
  \citenamefont {Mizera},\ and\ \citenamefont {Tourkine}}]{Casali:2020knc}%
  \BibitemOpen
  \bibfield  {author} {\bibinfo {author} {\bibfnamefont {E.}~\bibnamefont
  {Casali}}, \bibinfo {author} {\bibfnamefont {S.}~\bibnamefont {Mizera}},\
  and\ \bibinfo {author} {\bibfnamefont {P.}~\bibnamefont {Tourkine}},\
  }\bibfield  {title} {\bibinfo {title} {{Loop amplitudes monodromy relations
  and color-kinematics duality}},\ }\href
  {https://doi.org/10.1007/JHEP03(2021)048} {\bibfield  {journal} {\bibinfo
  {journal} {JHEP}\ }\textbf {\bibinfo {volume} {03}},\ \bibinfo {pages}
  {048}},\ \Eprint {https://arxiv.org/abs/2005.05329} {arXiv:2005.05329
  [hep-th]} \BibitemShut {NoStop}%
\bibitem [{\citenamefont {Borsten}\ \emph {et~al.}(2021)\citenamefont
  {Borsten}, \citenamefont {Kim}, \citenamefont {Jur\v{c}o}, \citenamefont
  {Macrelli}, \citenamefont {Saemann},\ and\ \citenamefont
  {Wolf}}]{Borsten:2021hua}%
  \BibitemOpen
  \bibfield  {author} {\bibinfo {author} {\bibfnamefont {L.}~\bibnamefont
  {Borsten}}, \bibinfo {author} {\bibfnamefont {H.}~\bibnamefont {Kim}},
  \bibinfo {author} {\bibfnamefont {B.}~\bibnamefont {Jur\v{c}o}}, \bibinfo
  {author} {\bibfnamefont {T.}~\bibnamefont {Macrelli}}, \bibinfo {author}
  {\bibfnamefont {C.}~\bibnamefont {Saemann}},\ and\ \bibinfo {author}
  {\bibfnamefont {M.}~\bibnamefont {Wolf}},\ }\bibfield  {title} {\bibinfo
  {title} {{Double Copy from Homotopy Algebras}},\ }\href@noop {} {\  (\bibinfo
  {year} {2021})},\ \Eprint {https://arxiv.org/abs/2102.11390}
  {arXiv:2102.11390 [hep-th]} \BibitemShut {NoStop}%
\bibitem [{\citenamefont {Bridges}\ and\ \citenamefont
  {Mafra}(2021)}]{Bridges:2021ebs}%
  \BibitemOpen
  \bibfield  {author} {\bibinfo {author} {\bibfnamefont {E.}~\bibnamefont
  {Bridges}}\ and\ \bibinfo {author} {\bibfnamefont {C.~R.}\ \bibnamefont
  {Mafra}},\ }\bibfield  {title} {\bibinfo {title} {{Local BCJ numerators for
  ten-dimensional SYM at one loop}},\ }\href@noop {} {\  (\bibinfo {year}
  {2021})},\ \Eprint {https://arxiv.org/abs/2102.12943} {arXiv:2102.12943
  [hep-th]} \BibitemShut {NoStop}%
\bibitem [{Note6()}]{Note6}%
  \BibitemOpen
  \bibinfo {note} {The tree-level (${\protect \mathcal J}^{(0)}=1$) version of
  this relation was revealed in string theory in \cite
  {Mafra:2011nv,Mafra:2011kj} and in the scattering equations (CHY) formalism
  in \cite {Cachazo:2013iea}; see also \cite {Azevedo:2018dgo,He:2018pol}. Our
  higher-loop formula is motivated by its verification at two loops in \cite
  {Geyer:2019hnn}. The higher-multiplicity extension is trivial.}\BibitemShut
  {Stop}%
\bibitem [{\citenamefont {Bjerrum-Bohr}\ \emph {et~al.}(2014)\citenamefont
  {Bjerrum-Bohr}, \citenamefont {Damgaard}, \citenamefont {Tourkine},\ and\
  \citenamefont {Vanhove}}]{Bjerrum-Bohr:2014qwa}%
  \BibitemOpen
  \bibfield  {author} {\bibinfo {author} {\bibfnamefont {N.~E.~J.}\
  \bibnamefont {Bjerrum-Bohr}}, \bibinfo {author} {\bibfnamefont {P.~H.}\
  \bibnamefont {Damgaard}}, \bibinfo {author} {\bibfnamefont {P.}~\bibnamefont
  {Tourkine}},\ and\ \bibinfo {author} {\bibfnamefont {P.}~\bibnamefont
  {Vanhove}},\ }\bibfield  {title} {\bibinfo {title} {{Scattering Equations and
  String Theory Amplitudes}},\ }\href
  {https://doi.org/10.1103/PhysRevD.90.106002} {\bibfield  {journal} {\bibinfo
  {journal} {Phys. Rev. D}\ }\textbf {\bibinfo {volume} {90}},\ \bibinfo
  {pages} {106002} (\bibinfo {year} {2014})},\ \Eprint
  {https://arxiv.org/abs/1403.4553} {arXiv:1403.4553 [hep-th]} \BibitemShut
  {NoStop}%
\bibitem [{\citenamefont {Mizera}(2020{\natexlab{b}})}]{Mizera:2019blq}%
  \BibitemOpen
  \bibfield  {author} {\bibinfo {author} {\bibfnamefont {S.}~\bibnamefont
  {Mizera}},\ }\bibfield  {title} {\bibinfo {title} {{Kinematic Jacobi Identity
  is a Residue Theorem: Geometry of Color-Kinematics Duality for Gauge and
  Gravity Amplitudes}},\ }\href
  {https://doi.org/10.1103/PhysRevLett.124.141601} {\bibfield  {journal}
  {\bibinfo  {journal} {Phys. Rev. Lett.}\ }\textbf {\bibinfo {volume} {124}},\
  \bibinfo {pages} {141601} (\bibinfo {year} {2020}{\natexlab{b}})},\ \Eprint
  {https://arxiv.org/abs/1912.03397} {arXiv:1912.03397 [hep-th]} \BibitemShut
  {NoStop}%
\bibitem [{\citenamefont
  {Kalyanapuram}(2021{\natexlab{a}})}]{Kalyanapuram:2021xow}%
  \BibitemOpen
  \bibfield  {author} {\bibinfo {author} {\bibfnamefont {N.}~\bibnamefont
  {Kalyanapuram}},\ }\bibfield  {title} {\bibinfo {title} {{Ambitwistor
  Integrands from Tensionless Chiral Superstring Integrands}},\ }\href@noop {}
  {\  (\bibinfo {year} {2021}{\natexlab{a}})},\ \Eprint
  {https://arxiv.org/abs/2103.07943} {arXiv:2103.07943 [hep-th]} \BibitemShut
  {NoStop}%
\bibitem [{\citenamefont
  {Kalyanapuram}(2021{\natexlab{b}})}]{Kalyanapuram:2021vjt}%
  \BibitemOpen
  \bibfield  {author} {\bibinfo {author} {\bibfnamefont {N.}~\bibnamefont
  {Kalyanapuram}},\ }\bibfield  {title} {\bibinfo {title} {{On Chiral Splitting
  and the Ambitwistor String}},\ }\href@noop {} {\  (\bibinfo {year}
  {2021}{\natexlab{b}})},\ \Eprint {https://arxiv.org/abs/2103.08584}
  {arXiv:2103.08584 [hep-th]} \BibitemShut {NoStop}%
\bibitem [{\citenamefont {Bern}\ \emph {et~al.}(1998)\citenamefont {Bern},
  \citenamefont {Dixon}, \citenamefont {Dunbar}, \citenamefont {Perelstein},\
  and\ \citenamefont {Rozowsky}}]{Bern:1998ug}%
  \BibitemOpen
  \bibfield  {author} {\bibinfo {author} {\bibfnamefont {Z.}~\bibnamefont
  {Bern}}, \bibinfo {author} {\bibfnamefont {L.~J.}\ \bibnamefont {Dixon}},
  \bibinfo {author} {\bibfnamefont {D.~C.}\ \bibnamefont {Dunbar}}, \bibinfo
  {author} {\bibfnamefont {M.}~\bibnamefont {Perelstein}},\ and\ \bibinfo
  {author} {\bibfnamefont {J.~S.}\ \bibnamefont {Rozowsky}},\ }\bibfield
  {title} {\bibinfo {title} {{On the relationship between Yang-Mills theory and
  gravity and its implication for ultraviolet divergences}},\ }\href
  {https://doi.org/10.1016/S0550-3213(98)00420-9} {\bibfield  {journal}
  {\bibinfo  {journal} {Nucl. Phys.}\ }\textbf {\bibinfo {volume} {B530}},\
  \bibinfo {pages} {401} (\bibinfo {year} {1998})},\ \Eprint
  {https://arxiv.org/abs/hep-th/9802162} {arXiv:hep-th/9802162 [hep-th]}
  \BibitemShut {NoStop}%
\bibitem [{Note7()}]{Note7}%
  \BibitemOpen
  \bibinfo {note} {Though the focus in \cite {Bern:1998ug} was 4D $\protect
  \mathcal N=8$ supergravity, the four-point loop integrand is valid in $D$
  dimensions. We use its 10D specialisation, with the appropriate prefactor
  ${\protect \mathcal R}^4$. The unregularised 10D supergravity loop {\protect
  \it integral} is UV divergent.}\BibitemShut {Stop}%
\bibitem [{Note8()}]{Note8}%
  \BibitemOpen
  \bibinfo {note} {They will be discussed elsewhere.}\BibitemShut {Stop}%
\bibitem [{Note9()}]{Note9}%
  \BibitemOpen
  \bibinfo {note} {The loop integrand in \cite {Bern:2010ue} was constructed
  using $D$-dimensional unitarity cuts. We use its 10D specialisation, with the
  appropriate prefactor ${\protect \mathcal R}^4$.}\BibitemShut {Stop}%
\bibitem [{Note10()}]{Note10}%
  \BibitemOpen
  \bibinfo {note} {Our convention for the external momenta is that they are
  incoming, whereas the convention in \cite {Bern:2010ue} was that they are
  outgoing. This affects the sign of the term linear in the loop
  momenta.}\BibitemShut {Stop}%
\bibitem [{Note11()}]{Note11}%
  \BibitemOpen
  \bibinfo {note} {The factor $(2\pi i)^{-3}$ in the second term of \protect
  \eqref {eq:Y1234}, introduced in the arXiv version v3, corrects a typo in
  earlier versions, including the version published in PRL.}\BibitemShut
  {Stop}%
\bibitem [{Note12()}]{Note12}%
  \BibitemOpen
  \bibinfo {note} {In summary, if a marked point $z_i$ is shifted by a
  `B-cycle', (i) the loop momentum associated to that cycle is shifted by
  $-k_i$ and (ii) the Abelian differential of the third kind has non-trivial
  monodromy. These two effects combine precisely to achieve homology
  invariance. See the very clear discussion for the two-loop five-point
  amplitude in \cite {DHoker:2020prr}, where the objects $g^I_{i,j}$ relate to
  $\omega _{3,4}(z_1)$ as $\omega _{3,4}(z_1)=(g^I_{1,3}-g^I_{1,4})\omega
  _I(z_1)$}\BibitemShut {NoStop}%
\bibitem [{\citenamefont {D'Hoker}\ and\ \citenamefont
  {Phong}(2002{\natexlab{a}})}]{DHoker:2002hof}%
  \BibitemOpen
  \bibfield  {author} {\bibinfo {author} {\bibfnamefont {E.}~\bibnamefont
  {D'Hoker}}\ and\ \bibinfo {author} {\bibfnamefont {D.~H.}\ \bibnamefont
  {Phong}},\ }\bibfield  {title} {\bibinfo {title} {{Lectures on two loop
  superstrings}},\ }\bibfield  {booktitle} {\emph {\bibinfo {booktitle}
  {{Superstring theory. Proceedings, International Conference, Hangzhou, P.R.
  China, August 12-15, 2002}}},\ }\href@noop {} {\bibfield  {journal} {\bibinfo
   {journal} {Conf. Proc.}\ }\textbf {\bibinfo {volume} {C0208124}},\ \bibinfo
  {pages} {85} (\bibinfo {year} {2002}{\natexlab{a}})},\ \bibinfo {note}
  {[,85(2002)]},\ \Eprint {https://arxiv.org/abs/hep-th/0211111}
  {arXiv:hep-th/0211111 [hep-th]} \BibitemShut {NoStop}%
\bibitem [{\citenamefont {D'Hoker}\ and\ \citenamefont
  {Phong}(2002{\natexlab{b}})}]{DHoker:2001kkt}%
  \BibitemOpen
  \bibfield  {author} {\bibinfo {author} {\bibfnamefont {E.}~\bibnamefont
  {D'Hoker}}\ and\ \bibinfo {author} {\bibfnamefont {D.~H.}\ \bibnamefont
  {Phong}},\ }\bibfield  {title} {\bibinfo {title} {{Two loop superstrings. 1.
  Main formulas}},\ }\href {https://doi.org/10.1016/S0370-2693(02)01255-8}
  {\bibfield  {journal} {\bibinfo  {journal} {Phys. Lett.}\ }\textbf {\bibinfo
  {volume} {B529}},\ \bibinfo {pages} {241} (\bibinfo {year}
  {2002}{\natexlab{b}})},\ \Eprint {https://arxiv.org/abs/hep-th/0110247}
  {arXiv:hep-th/0110247 [hep-th]} \BibitemShut {NoStop}%
\bibitem [{\citenamefont {D'Hoker}\ and\ \citenamefont
  {Phong}(2002{\natexlab{c}})}]{DHoker:2001qqx}%
  \BibitemOpen
  \bibfield  {author} {\bibinfo {author} {\bibfnamefont {E.}~\bibnamefont
  {D'Hoker}}\ and\ \bibinfo {author} {\bibfnamefont {D.~H.}\ \bibnamefont
  {Phong}},\ }\bibfield  {title} {\bibinfo {title} {{Two loop superstrings. 2.
  The Chiral measure on moduli space}},\ }\href
  {https://doi.org/10.1016/S0550-3213(02)00431-5} {\bibfield  {journal}
  {\bibinfo  {journal} {Nucl. Phys.}\ }\textbf {\bibinfo {volume} {B636}},\
  \bibinfo {pages} {3} (\bibinfo {year} {2002}{\natexlab{c}})},\ \Eprint
  {https://arxiv.org/abs/hep-th/0110283} {arXiv:hep-th/0110283 [hep-th]}
  \BibitemShut {NoStop}%
\bibitem [{\citenamefont {D'Hoker}\ and\ \citenamefont
  {Phong}(2002{\natexlab{d}})}]{DHoker:2001foj}%
  \BibitemOpen
  \bibfield  {author} {\bibinfo {author} {\bibfnamefont {E.}~\bibnamefont
  {D'Hoker}}\ and\ \bibinfo {author} {\bibfnamefont {D.~H.}\ \bibnamefont
  {Phong}},\ }\bibfield  {title} {\bibinfo {title} {{Two loop superstrings. 3.
  Slice independence and absence of ambiguities}},\ }\href
  {https://doi.org/10.1016/S0550-3213(02)00432-7} {\bibfield  {journal}
  {\bibinfo  {journal} {Nucl. Phys.}\ }\textbf {\bibinfo {volume} {B636}},\
  \bibinfo {pages} {61} (\bibinfo {year} {2002}{\natexlab{d}})},\ \Eprint
  {https://arxiv.org/abs/hep-th/0111016} {arXiv:hep-th/0111016 [hep-th]}
  \BibitemShut {NoStop}%
\bibitem [{\citenamefont {D'Hoker}\ and\ \citenamefont
  {Phong}(2002{\natexlab{e}})}]{DHoker:2001jaf}%
  \BibitemOpen
  \bibfield  {author} {\bibinfo {author} {\bibfnamefont {E.}~\bibnamefont
  {D'Hoker}}\ and\ \bibinfo {author} {\bibfnamefont {D.~H.}\ \bibnamefont
  {Phong}},\ }\bibfield  {title} {\bibinfo {title} {{Two loop superstrings 4:
  The Cosmological constant and modular forms}},\ }\href
  {https://doi.org/10.1016/S0550-3213(02)00516-3} {\bibfield  {journal}
  {\bibinfo  {journal} {Nucl. Phys.}\ }\textbf {\bibinfo {volume} {B639}},\
  \bibinfo {pages} {129} (\bibinfo {year} {2002}{\natexlab{e}})},\ \Eprint
  {https://arxiv.org/abs/hep-th/0111040} {arXiv:hep-th/0111040 [hep-th]}
  \BibitemShut {NoStop}%
\bibitem [{\citenamefont {D'Hoker}\ and\ \citenamefont
  {Phong}(2005{\natexlab{d}})}]{DHoker:2005dys}%
  \BibitemOpen
  \bibfield  {author} {\bibinfo {author} {\bibfnamefont {E.}~\bibnamefont
  {D'Hoker}}\ and\ \bibinfo {author} {\bibfnamefont {D.~H.}\ \bibnamefont
  {Phong}},\ }\bibfield  {title} {\bibinfo {title} {{Two-loop superstrings. V.
  Gauge slice independence of the N-point function}},\ }\href
  {https://doi.org/10.1016/j.nuclphysb.2005.02.042} {\bibfield  {journal}
  {\bibinfo  {journal} {Nucl. Phys.}\ }\textbf {\bibinfo {volume} {B715}},\
  \bibinfo {pages} {91} (\bibinfo {year} {2005}{\natexlab{d}})},\ \Eprint
  {https://arxiv.org/abs/hep-th/0501196} {arXiv:hep-th/0501196 [hep-th]}
  \BibitemShut {NoStop}%
\bibitem [{\citenamefont {Tsuyumine}(1986)}]{tsuyumine1986}%
  \BibitemOpen
  \bibfield  {author} {\bibinfo {author} {\bibfnamefont {S.}~\bibnamefont
  {Tsuyumine}},\ }\bibfield  {title} {\bibinfo {title} {{On Siegel modular
  forms of degree three}},\ }\href@noop {} {\bibfield  {journal} {\bibinfo
  {journal} {Amer. J. Math.}\ }\textbf {\bibinfo {volume} {108}},\ \bibinfo
  {pages} {755} (\bibinfo {year} {1986})}\BibitemShut {NoStop}%
\bibitem [{\citenamefont {Witten}(2015)}]{Witten:2015hwa}%
  \BibitemOpen
  \bibfield  {author} {\bibinfo {author} {\bibfnamefont {E.}~\bibnamefont
  {Witten}},\ }\bibfield  {title} {\bibinfo {title} {{The Super Period Matrix
  With Ramond Punctures}},\ }\href
  {https://doi.org/10.1016/j.geomphys.2015.02.017} {\bibfield  {journal}
  {\bibinfo  {journal} {J. Geom. Phys.}\ }\textbf {\bibinfo {volume} {92}},\
  \bibinfo {pages} {210} (\bibinfo {year} {2015})},\ \Eprint
  {https://arxiv.org/abs/1501.02499} {arXiv:1501.02499 [hep-th]} \BibitemShut
  {NoStop}%
\bibitem [{Note13()}]{Note13}%
  \BibitemOpen
  \bibinfo {note} {And the non-vanishing of another modular form called $\Sigma
  _{140}$ in the classical reference \cite {Igusa}.}\BibitemShut {Stop}%
\bibitem [{Note14()}]{Note14}%
  \BibitemOpen
  \bibinfo {note} {One may also ask why a sum with $$ S_\delta (z_1,z_3)^2
  S_\delta (z_2,z_4)^2 + S_\delta (z_1,z_4)^2 S_\delta (z_2,z_3)^2 $$ is absent
  from our result, since it has the correct symmetries. It turns out that this
  sum gives precisely twice the sum \protect \eqref {eq:Szegob}.}\BibitemShut
  {Stop}%
\bibitem [{\citenamefont {Grushevsky}\ and\ \citenamefont
  {Salvati~Manni}(2010)}]{Grushevsky:2008qp}%
  \BibitemOpen
  \bibfield  {author} {\bibinfo {author} {\bibfnamefont {S.}~\bibnamefont
  {Grushevsky}}\ and\ \bibinfo {author} {\bibfnamefont {R.}~\bibnamefont
  {Salvati~Manni}},\ }\bibfield  {title} {\bibinfo {title} {{The Vanishing of
  two-point functions for three-loop superstring scattering amplitudes}},\
  }\href {https://doi.org/10.1007/s00220-009-0967-1} {\bibfield  {journal}
  {\bibinfo  {journal} {Commun. Math. Phys.}\ }\textbf {\bibinfo {volume}
  {294}},\ \bibinfo {pages} {343} (\bibinfo {year} {2010})},\ \Eprint
  {https://arxiv.org/abs/0806.0354} {arXiv:0806.0354 [hep-th]} \BibitemShut
  {NoStop}%
\bibitem [{\citenamefont {Matone}\ and\ \citenamefont
  {Volpato}(2009)}]{Matone:2008td}%
  \BibitemOpen
  \bibfield  {author} {\bibinfo {author} {\bibfnamefont {M.}~\bibnamefont
  {Matone}}\ and\ \bibinfo {author} {\bibfnamefont {R.}~\bibnamefont
  {Volpato}},\ }\bibfield  {title} {\bibinfo {title} {{Superstring measure and
  non-renormalization of the three-point amplitude}},\ }\href
  {https://doi.org/10.1016/j.nuclphysb.2008.08.011} {\bibfield  {journal}
  {\bibinfo  {journal} {Nucl. Phys. B}\ }\textbf {\bibinfo {volume} {806}},\
  \bibinfo {pages} {735} (\bibinfo {year} {2009})},\ \Eprint
  {https://arxiv.org/abs/0806.4370} {arXiv:0806.4370 [hep-th]} \BibitemShut
  {NoStop}%
\bibitem [{\citenamefont {Green}\ and\ \citenamefont
  {Vanhove}(2006)}]{Green:2005ba}%
  \BibitemOpen
  \bibfield  {author} {\bibinfo {author} {\bibfnamefont {M.~B.}\ \bibnamefont
  {Green}}\ and\ \bibinfo {author} {\bibfnamefont {P.}~\bibnamefont
  {Vanhove}},\ }\bibfield  {title} {\bibinfo {title} {{Duality and higher
  derivative terms in M theory}},\ }\href
  {https://doi.org/10.1088/1126-6708/2006/01/093} {\bibfield  {journal}
  {\bibinfo  {journal} {JHEP}\ }\textbf {\bibinfo {volume} {01}},\ \bibinfo
  {pages} {093}},\ \Eprint {https://arxiv.org/abs/hep-th/0510027}
  {arXiv:hep-th/0510027} \BibitemShut {NoStop}%
\bibitem [{\citenamefont {Berkovits}(2000)}]{Berkovits:2000fe}%
  \BibitemOpen
  \bibfield  {author} {\bibinfo {author} {\bibfnamefont {N.}~\bibnamefont
  {Berkovits}},\ }\bibfield  {title} {\bibinfo {title} {{Super Poincare
  covariant quantization of the superstring}},\ }\href
  {https://doi.org/10.1088/1126-6708/2000/04/018} {\bibfield  {journal}
  {\bibinfo  {journal} {JHEP}\ }\textbf {\bibinfo {volume} {04}},\ \bibinfo
  {pages} {018}},\ \Eprint {https://arxiv.org/abs/hep-th/0001035}
  {arXiv:hep-th/0001035} \BibitemShut {NoStop}%
\bibitem [{\citenamefont {Berkovits}(2001)}]{Berkovits:2001rb}%
  \BibitemOpen
  \bibfield  {author} {\bibinfo {author} {\bibfnamefont {N.}~\bibnamefont
  {Berkovits}},\ }\bibfield  {title} {\bibinfo {title} {{Covariant quantization
  of the superparticle using pure spinors}},\ }\href
  {https://doi.org/10.1088/1126-6708/2001/09/016} {\bibfield  {journal}
  {\bibinfo  {journal} {JHEP}\ }\textbf {\bibinfo {volume} {09}},\ \bibinfo
  {pages} {016}},\ \Eprint {https://arxiv.org/abs/hep-th/0105050}
  {arXiv:hep-th/0105050} \BibitemShut {NoStop}%
\bibitem [{\citenamefont {Berkovits}(2003)}]{Berkovits:2002zk}%
  \BibitemOpen
  \bibfield  {author} {\bibinfo {author} {\bibfnamefont {N.}~\bibnamefont
  {Berkovits}},\ }\bibfield  {title} {\bibinfo {title} {{ICTP lectures on
  covariant quantization of the superstring}},\ }\href@noop {} {\bibfield
  {journal} {\bibinfo  {journal} {ICTP Lect. Notes Ser.}\ }\textbf {\bibinfo
  {volume} {13}},\ \bibinfo {pages} {57} (\bibinfo {year} {2003})},\ \Eprint
  {https://arxiv.org/abs/hep-th/0209059} {arXiv:hep-th/0209059} \BibitemShut
  {NoStop}%
\bibitem [{\citenamefont {Berkovits}(2004)}]{Berkovits:2004px}%
  \BibitemOpen
  \bibfield  {author} {\bibinfo {author} {\bibfnamefont {N.}~\bibnamefont
  {Berkovits}},\ }\bibfield  {title} {\bibinfo {title} {{Multiloop amplitudes
  and vanishing theorems using the pure spinor formalism for the
  superstring}},\ }\href {https://doi.org/10.1088/1126-6708/2004/09/047}
  {\bibfield  {journal} {\bibinfo  {journal} {JHEP}\ }\textbf {\bibinfo
  {volume} {09}},\ \bibinfo {pages} {047}},\ \Eprint
  {https://arxiv.org/abs/hep-th/0406055} {arXiv:hep-th/0406055} \BibitemShut
  {NoStop}%
\bibitem [{\citenamefont {D'Hoker}\ \emph {et~al.}(2020)\citenamefont
  {D'Hoker}, \citenamefont {Mafra}, \citenamefont {Pioline},\ and\
  \citenamefont {Schlotterer}}]{DHoker:2020prr}%
  \BibitemOpen
  \bibfield  {author} {\bibinfo {author} {\bibfnamefont {E.}~\bibnamefont
  {D'Hoker}}, \bibinfo {author} {\bibfnamefont {C.~R.}\ \bibnamefont {Mafra}},
  \bibinfo {author} {\bibfnamefont {B.}~\bibnamefont {Pioline}},\ and\ \bibinfo
  {author} {\bibfnamefont {O.}~\bibnamefont {Schlotterer}},\ }\bibfield
  {title} {\bibinfo {title} {{Two-loop superstring five-point amplitudes. Part
  I. Construction via chiral splitting and pure spinors}},\ }\href
  {https://doi.org/10.1007/JHEP08(2020)135} {\bibfield  {journal} {\bibinfo
  {journal} {JHEP}\ }\textbf {\bibinfo {volume} {08}},\ \bibinfo {pages}
  {135}},\ \Eprint {https://arxiv.org/abs/2006.05270} {arXiv:2006.05270
  [hep-th]} \BibitemShut {NoStop}%
\bibitem [{\citenamefont {D'Hoker}\ \emph {et~al.}(2005)\citenamefont
  {D'Hoker}, \citenamefont {Gutperle},\ and\ \citenamefont
  {Phong}}]{DHoker:2005jhf}%
  \BibitemOpen
  \bibfield  {author} {\bibinfo {author} {\bibfnamefont {E.}~\bibnamefont
  {D'Hoker}}, \bibinfo {author} {\bibfnamefont {M.}~\bibnamefont {Gutperle}},\
  and\ \bibinfo {author} {\bibfnamefont {D.~H.}\ \bibnamefont {Phong}},\
  }\bibfield  {title} {\bibinfo {title} {{Two-loop superstrings and
  S-duality}},\ }\href {https://doi.org/10.1016/j.nuclphysb.2005.06.010}
  {\bibfield  {journal} {\bibinfo  {journal} {Nucl. Phys.}\ }\textbf {\bibinfo
  {volume} {B722}},\ \bibinfo {pages} {81} (\bibinfo {year} {2005})},\ \Eprint
  {https://arxiv.org/abs/hep-th/0503180} {arXiv:hep-th/0503180 [hep-th]}
  \BibitemShut {NoStop}%
\bibitem [{\citenamefont {D'Hoker}\ and\ \citenamefont
  {Green}(2013)}]{DHoker:2013fcx}%
  \BibitemOpen
  \bibfield  {author} {\bibinfo {author} {\bibfnamefont {E.}~\bibnamefont
  {D'Hoker}}\ and\ \bibinfo {author} {\bibfnamefont {M.~B.}\ \bibnamefont
  {Green}},\ }\bibfield  {title} {\bibinfo {title} {{Zhang-Kawazumi Invariants
  and Superstring Amplitudes}},\ }\href@noop {} {\  (\bibinfo {year} {2013})},\
  \Eprint {https://arxiv.org/abs/1308.4597} {arXiv:1308.4597 [hep-th]}
  \BibitemShut {NoStop}%
\bibitem [{\citenamefont {D'Hoker}\ \emph {et~al.}(2015)\citenamefont
  {D'Hoker}, \citenamefont {Green}, \citenamefont {Pioline},\ and\
  \citenamefont {Russo}}]{DHoker:2014oxd}%
  \BibitemOpen
  \bibfield  {author} {\bibinfo {author} {\bibfnamefont {E.}~\bibnamefont
  {D'Hoker}}, \bibinfo {author} {\bibfnamefont {M.~B.}\ \bibnamefont {Green}},
  \bibinfo {author} {\bibfnamefont {B.}~\bibnamefont {Pioline}},\ and\ \bibinfo
  {author} {\bibfnamefont {R.}~\bibnamefont {Russo}},\ }\bibfield  {title}
  {\bibinfo {title} {{Matching the $D^{6}R^{4}$ interaction at two-loops}},\
  }\href {https://doi.org/10.1007/JHEP01(2015)031} {\bibfield  {journal}
  {\bibinfo  {journal} {JHEP}\ }\textbf {\bibinfo {volume} {01}},\ \bibinfo
  {pages} {031}},\ \Eprint {https://arxiv.org/abs/1405.6226} {arXiv:1405.6226
  [hep-th]} \BibitemShut {NoStop}%
\bibitem [{\citenamefont {D'Hoker}\ \emph {et~al.}(2021)\citenamefont
  {D'Hoker}, \citenamefont {Mafra}, \citenamefont {Pioline},\ and\
  \citenamefont {Schlotterer}}]{DHoker:2020tcq}%
  \BibitemOpen
  \bibfield  {author} {\bibinfo {author} {\bibfnamefont {E.}~\bibnamefont
  {D'Hoker}}, \bibinfo {author} {\bibfnamefont {C.~R.}\ \bibnamefont {Mafra}},
  \bibinfo {author} {\bibfnamefont {B.}~\bibnamefont {Pioline}},\ and\ \bibinfo
  {author} {\bibfnamefont {O.}~\bibnamefont {Schlotterer}},\ }\bibfield
  {title} {\bibinfo {title} {{Two-loop superstring five-point amplitudes. Part
  II. Low energy expansion and S-duality}},\ }\href
  {https://doi.org/10.1007/JHEP02(2021)139} {\bibfield  {journal} {\bibinfo
  {journal} {JHEP}\ }\textbf {\bibinfo {volume} {02}},\ \bibinfo {pages}
  {139}},\ \Eprint {https://arxiv.org/abs/2008.08687} {arXiv:2008.08687
  [hep-th]} \BibitemShut {NoStop}%
\bibitem [{\citenamefont {Guerrieri}\ \emph {et~al.}(2021)\citenamefont
  {Guerrieri}, \citenamefont {Penedones},\ and\ \citenamefont
  {Vieira}}]{Guerrieri:2021ivu}%
  \BibitemOpen
  \bibfield  {author} {\bibinfo {author} {\bibfnamefont {A.}~\bibnamefont
  {Guerrieri}}, \bibinfo {author} {\bibfnamefont {J.}~\bibnamefont
  {Penedones}},\ and\ \bibinfo {author} {\bibfnamefont {P.}~\bibnamefont
  {Vieira}},\ }\bibfield  {title} {\bibinfo {title} {{Where is String
  Theory?}},\ }\href@noop {} {\  (\bibinfo {year} {2021})},\ \Eprint
  {https://arxiv.org/abs/2102.02847} {arXiv:2102.02847 [hep-th]} \BibitemShut
  {NoStop}%
\bibitem [{\citenamefont {Tsuchiya}(1989)}]{Tsuchiya:1988va}%
  \BibitemOpen
  \bibfield  {author} {\bibinfo {author} {\bibfnamefont {A.}~\bibnamefont
  {Tsuchiya}},\ }\bibfield  {title} {\bibinfo {title} {{More on One Loop
  Massless Amplitudes of Superstring Theories}},\ }\href
  {https://doi.org/10.1103/PhysRevD.39.1626} {\bibfield  {journal} {\bibinfo
  {journal} {Phys. Rev.}\ }\textbf {\bibinfo {volume} {D39}},\ \bibinfo {pages}
  {1626} (\bibinfo {year} {1989})}\BibitemShut {NoStop}%
\bibitem [{\citenamefont {Richards}(2008)}]{Richards:2008jg}%
  \BibitemOpen
  \bibfield  {author} {\bibinfo {author} {\bibfnamefont {D.~M.}\ \bibnamefont
  {Richards}},\ }\bibfield  {title} {\bibinfo {title} {{The One-Loop
  Five-Graviton Amplitude and the Effective Action}},\ }\href
  {https://doi.org/10.1088/1126-6708/2008/10/042} {\bibfield  {journal}
  {\bibinfo  {journal} {JHEP}\ }\textbf {\bibinfo {volume} {10}},\ \bibinfo
  {pages} {042}},\ \Eprint {https://arxiv.org/abs/0807.2421} {arXiv:0807.2421
  [hep-th]} \BibitemShut {NoStop}%
\bibitem [{\citenamefont {Tsuchiya}(2012)}]{Tsuchiya:2012nf}%
  \BibitemOpen
  \bibfield  {author} {\bibinfo {author} {\bibfnamefont {A.~G.}\ \bibnamefont
  {Tsuchiya}},\ }\bibfield  {title} {\bibinfo {title} {{On the pole structures
  of the disconnected part of hyper elliptic g loop M point super string
  amplitudes}},\ }\href@noop {} {\  (\bibinfo {year} {2012})},\ \Eprint
  {https://arxiv.org/abs/1209.6117} {arXiv:1209.6117 [hep-th]} \BibitemShut
  {NoStop}%
\bibitem [{\citenamefont {Green}\ \emph {et~al.}(2013)\citenamefont {Green},
  \citenamefont {Mafra},\ and\ \citenamefont {Schlotterer}}]{Green:2013bza}%
  \BibitemOpen
  \bibfield  {author} {\bibinfo {author} {\bibfnamefont {M.~B.}\ \bibnamefont
  {Green}}, \bibinfo {author} {\bibfnamefont {C.~R.}\ \bibnamefont {Mafra}},\
  and\ \bibinfo {author} {\bibfnamefont {O.}~\bibnamefont {Schlotterer}},\
  }\bibfield  {title} {\bibinfo {title} {{Multiparticle one-loop amplitudes and
  S-duality in closed superstring theory}},\ }\href
  {https://doi.org/10.1007/JHEP10(2013)188} {\bibfield  {journal} {\bibinfo
  {journal} {JHEP}\ }\textbf {\bibinfo {volume} {10}},\ \bibinfo {pages}
  {188}},\ \Eprint {https://arxiv.org/abs/1307.3534} {arXiv:1307.3534 [hep-th]}
  \BibitemShut {NoStop}%
\bibitem [{\citenamefont {Mafra}\ and\ \citenamefont
  {Schlotterer}(2016)}]{Mafra:2016nwr}%
  \BibitemOpen
  \bibfield  {author} {\bibinfo {author} {\bibfnamefont {C.~R.}\ \bibnamefont
  {Mafra}}\ and\ \bibinfo {author} {\bibfnamefont {O.}~\bibnamefont
  {Schlotterer}},\ }\bibfield  {title} {\bibinfo {title} {{One-loop superstring
  six-point amplitudes and anomalies in pure spinor superspace}},\ }\href
  {https://doi.org/10.1007/JHEP04(2016)148} {\bibfield  {journal} {\bibinfo
  {journal} {JHEP}\ }\textbf {\bibinfo {volume} {04}},\ \bibinfo {pages}
  {148}},\ \Eprint {https://arxiv.org/abs/1603.04790} {arXiv:1603.04790
  [hep-th]} \BibitemShut {NoStop}%
\bibitem [{\citenamefont {Mafra}\ and\ \citenamefont
  {Schlotterer}(2019{\natexlab{a}})}]{Mafra:2018nla}%
  \BibitemOpen
  \bibfield  {author} {\bibinfo {author} {\bibfnamefont {C.~R.}\ \bibnamefont
  {Mafra}}\ and\ \bibinfo {author} {\bibfnamefont {O.}~\bibnamefont
  {Schlotterer}},\ }\bibfield  {title} {\bibinfo {title} {{Towards the n-point
  one-loop superstring amplitude. Part I. Pure spinors and superfield
  kinematics}},\ }\href {https://doi.org/10.1007/JHEP08(2019)090} {\bibfield
  {journal} {\bibinfo  {journal} {JHEP}\ }\textbf {\bibinfo {volume} {08}},\
  \bibinfo {pages} {090}},\ \Eprint {https://arxiv.org/abs/1812.10969}
  {arXiv:1812.10969 [hep-th]} \BibitemShut {NoStop}%
\bibitem [{\citenamefont {Mafra}\ and\ \citenamefont
  {Schlotterer}(2019{\natexlab{b}})}]{Mafra:2018pll}%
  \BibitemOpen
  \bibfield  {author} {\bibinfo {author} {\bibfnamefont {C.~R.}\ \bibnamefont
  {Mafra}}\ and\ \bibinfo {author} {\bibfnamefont {O.}~\bibnamefont
  {Schlotterer}},\ }\bibfield  {title} {\bibinfo {title} {{Towards the n-point
  one-loop superstring amplitude. Part II. Worldsheet functions and their
  duality to kinematics}},\ }\href {https://doi.org/10.1007/JHEP08(2019)091}
  {\bibfield  {journal} {\bibinfo  {journal} {JHEP}\ }\textbf {\bibinfo
  {volume} {08}},\ \bibinfo {pages} {091}},\ \Eprint
  {https://arxiv.org/abs/1812.10970} {arXiv:1812.10970 [hep-th]} \BibitemShut
  {NoStop}%
\bibitem [{\citenamefont {Mafra}\ and\ \citenamefont
  {Schlotterer}(2019{\natexlab{c}})}]{Mafra:2018qqe}%
  \BibitemOpen
  \bibfield  {author} {\bibinfo {author} {\bibfnamefont {C.~R.}\ \bibnamefont
  {Mafra}}\ and\ \bibinfo {author} {\bibfnamefont {O.}~\bibnamefont
  {Schlotterer}},\ }\bibfield  {title} {\bibinfo {title} {{Towards the n-point
  one-loop superstring amplitude. Part III. One-loop correlators and their
  double-copy structure}},\ }\href {https://doi.org/10.1007/JHEP08(2019)092}
  {\bibfield  {journal} {\bibinfo  {journal} {JHEP}\ }\textbf {\bibinfo
  {volume} {08}},\ \bibinfo {pages} {092}},\ \Eprint
  {https://arxiv.org/abs/1812.10971} {arXiv:1812.10971 [hep-th]} \BibitemShut
  {NoStop}%
\bibitem [{\citenamefont {Grushevsky}(2009)}]{Grushevsky:2008zm}%
  \BibitemOpen
  \bibfield  {author} {\bibinfo {author} {\bibfnamefont {S.}~\bibnamefont
  {Grushevsky}},\ }\bibfield  {title} {\bibinfo {title} {{Superstring
  scattering amplitudes in higher genus}},\ }\href
  {https://doi.org/10.1007/s00220-008-0635-x} {\bibfield  {journal} {\bibinfo
  {journal} {Commun. Math. Phys.}\ }\textbf {\bibinfo {volume} {287}},\
  \bibinfo {pages} {749} (\bibinfo {year} {2009})},\ \Eprint
  {https://arxiv.org/abs/0803.3469} {arXiv:0803.3469 [hep-th]} \BibitemShut
  {NoStop}%
\bibitem [{\citenamefont {Cacciatori}\ \emph
  {et~al.}(2008{\natexlab{b}})\citenamefont {Cacciatori}, \citenamefont
  {Dalla~Piazza},\ and\ \citenamefont {van Geemen}}]{Cacciatori:2008pj}%
  \BibitemOpen
  \bibfield  {author} {\bibinfo {author} {\bibfnamefont {S.~L.}\ \bibnamefont
  {Cacciatori}}, \bibinfo {author} {\bibfnamefont {F.}~\bibnamefont
  {Dalla~Piazza}},\ and\ \bibinfo {author} {\bibfnamefont {B.}~\bibnamefont
  {van Geemen}},\ }\bibfield  {title} {\bibinfo {title} {{Genus four
  superstring measures}},\ }\href {https://doi.org/10.1007/s11005-008-0260-9}
  {\bibfield  {journal} {\bibinfo  {journal} {Lett. Math. Phys.}\ }\textbf
  {\bibinfo {volume} {85}},\ \bibinfo {pages} {185} (\bibinfo {year}
  {2008}{\natexlab{b}})},\ \Eprint {https://arxiv.org/abs/0804.0457}
  {arXiv:0804.0457 [hep-th]} \BibitemShut {NoStop}%
\bibitem [{\citenamefont {Salvati-Manni}(2008)}]{SalvatiManni:2008qa}%
  \BibitemOpen
  \bibfield  {author} {\bibinfo {author} {\bibfnamefont {R.}~\bibnamefont
  {Salvati-Manni}},\ }\bibfield  {title} {\bibinfo {title} {{Remarks on
  Superstring amplitudes in higher genus}},\ }\href
  {https://doi.org/10.1016/j.nuclphysb.2008.05.009} {\bibfield  {journal}
  {\bibinfo  {journal} {Nucl. Phys. B}\ }\textbf {\bibinfo {volume} {801}},\
  \bibinfo {pages} {163} (\bibinfo {year} {2008})},\ \Eprint
  {https://arxiv.org/abs/0804.0512} {arXiv:0804.0512 [hep-th]} \BibitemShut
  {NoStop}%
\bibitem [{\citenamefont {Morozov}(2008)}]{Morozov:2008wz}%
  \BibitemOpen
  \bibfield  {author} {\bibinfo {author} {\bibfnamefont {A.}~\bibnamefont
  {Morozov}},\ }\bibfield  {title} {\bibinfo {title} {{NSR Superstring Measures
  Revisited}},\ }\href {https://doi.org/10.1088/1126-6708/2008/05/086}
  {\bibfield  {journal} {\bibinfo  {journal} {JHEP}\ }\textbf {\bibinfo
  {volume} {05}},\ \bibinfo {pages} {086}},\ \Eprint
  {https://arxiv.org/abs/0804.3167} {arXiv:0804.3167 [hep-th]} \BibitemShut
  {NoStop}%
\bibitem [{\citenamefont {Grushevsky}\ and\ \citenamefont
  {Salvati~Manni}(2011)}]{Grushevsky:2008zp}%
  \BibitemOpen
  \bibfield  {author} {\bibinfo {author} {\bibfnamefont {S.}~\bibnamefont
  {Grushevsky}}\ and\ \bibinfo {author} {\bibfnamefont {R.}~\bibnamefont
  {Salvati~Manni}},\ }\bibfield  {title} {\bibinfo {title} {{The superstring
  cosmological constant and the Schottky form in genus 5}},\ }\href
  {https://doi.org/10.1353/ajm.2011.0028} {\bibfield  {journal} {\bibinfo
  {journal} {Am. J. Math.}\ }\textbf {\bibinfo {volume} {133}},\ \bibinfo
  {pages} {1007} (\bibinfo {year} {2011})},\ \Eprint
  {https://arxiv.org/abs/0809.1391} {arXiv:0809.1391 [math.AG]} \BibitemShut
  {NoStop}%
\bibitem [{\citenamefont {Matone}\ and\ \citenamefont
  {Volpato}(2010)}]{Matone:2010yv}%
  \BibitemOpen
  \bibfield  {author} {\bibinfo {author} {\bibfnamefont {M.}~\bibnamefont
  {Matone}}\ and\ \bibinfo {author} {\bibfnamefont {R.}~\bibnamefont
  {Volpato}},\ }\bibfield  {title} {\bibinfo {title} {{Getting superstring
  amplitudes by degenerating Riemann surfaces}},\ }\href
  {https://doi.org/10.1016/j.nuclphysb.2010.05.020} {\bibfield  {journal}
  {\bibinfo  {journal} {Nucl. Phys. B}\ }\textbf {\bibinfo {volume} {839}},\
  \bibinfo {pages} {21} (\bibinfo {year} {2010})},\ \Eprint
  {https://arxiv.org/abs/1003.3452} {arXiv:1003.3452 [hep-th]} \BibitemShut
  {NoStop}%
\bibitem [{\citenamefont {Matone}\ and\ \citenamefont
  {Volpato}(2006)}]{Matone:2005vm}%
  \BibitemOpen
  \bibfield  {author} {\bibinfo {author} {\bibfnamefont {M.}~\bibnamefont
  {Matone}}\ and\ \bibinfo {author} {\bibfnamefont {R.}~\bibnamefont
  {Volpato}},\ }\bibfield  {title} {\bibinfo {title} {{Higher genus superstring
  amplitudes from the geometry of moduli space}},\ }\href
  {https://doi.org/10.1016/j.nuclphysb.2005.10.036} {\bibfield  {journal}
  {\bibinfo  {journal} {Nucl. Phys. B}\ }\textbf {\bibinfo {volume} {732}},\
  \bibinfo {pages} {321} (\bibinfo {year} {2006})},\ \Eprint
  {https://arxiv.org/abs/hep-th/0506231} {arXiv:hep-th/0506231} \BibitemShut
  {NoStop}%
\bibitem [{\citenamefont {Bern}\ \emph {et~al.}(2012)\citenamefont {Bern},
  \citenamefont {Carrasco}, \citenamefont {Dixon}, \citenamefont {Johansson},\
  and\ \citenamefont {Roiban}}]{Bern:2012uf}%
  \BibitemOpen
  \bibfield  {author} {\bibinfo {author} {\bibfnamefont {Z.}~\bibnamefont
  {Bern}}, \bibinfo {author} {\bibfnamefont {J.~J.~M.}\ \bibnamefont
  {Carrasco}}, \bibinfo {author} {\bibfnamefont {L.~J.}\ \bibnamefont {Dixon}},
  \bibinfo {author} {\bibfnamefont {H.}~\bibnamefont {Johansson}},\ and\
  \bibinfo {author} {\bibfnamefont {R.}~\bibnamefont {Roiban}},\ }\bibfield
  {title} {\bibinfo {title} {{Simplifying Multiloop Integrands and Ultraviolet
  Divergences of Gauge Theory and Gravity Amplitudes}},\ }\href
  {https://doi.org/10.1103/PhysRevD.85.105014} {\bibfield  {journal} {\bibinfo
  {journal} {Phys. Rev. D}\ }\textbf {\bibinfo {volume} {85}},\ \bibinfo
  {pages} {105014} (\bibinfo {year} {2012})},\ \Eprint
  {https://arxiv.org/abs/1201.5366} {arXiv:1201.5366 [hep-th]} \BibitemShut
  {NoStop}%
\bibitem [{\citenamefont {Bern}\ \emph
  {et~al.}(2017{\natexlab{a}})\citenamefont {Bern}, \citenamefont {Carrasco},
  \citenamefont {Chen}, \citenamefont {Johansson},\ and\ \citenamefont
  {Roiban}}]{Bern:2017yxu}%
  \BibitemOpen
  \bibfield  {author} {\bibinfo {author} {\bibfnamefont {Z.}~\bibnamefont
  {Bern}}, \bibinfo {author} {\bibfnamefont {J.~J.}\ \bibnamefont {Carrasco}},
  \bibinfo {author} {\bibfnamefont {W.-M.}\ \bibnamefont {Chen}}, \bibinfo
  {author} {\bibfnamefont {H.}~\bibnamefont {Johansson}},\ and\ \bibinfo
  {author} {\bibfnamefont {R.}~\bibnamefont {Roiban}},\ }\bibfield  {title}
  {\bibinfo {title} {{Gravity Amplitudes as Generalized Double Copies of
  Gauge-Theory Amplitudes}},\ }\href
  {https://doi.org/10.1103/PhysRevLett.118.181602} {\bibfield  {journal}
  {\bibinfo  {journal} {Phys. Rev. Lett.}\ }\textbf {\bibinfo {volume} {118}},\
  \bibinfo {pages} {181602} (\bibinfo {year} {2017}{\natexlab{a}})},\ \Eprint
  {https://arxiv.org/abs/1701.02519} {arXiv:1701.02519 [hep-th]} \BibitemShut
  {NoStop}%
\bibitem [{\citenamefont {Bern}\ \emph
  {et~al.}(2017{\natexlab{b}})\citenamefont {Bern}, \citenamefont {Carrasco},
  \citenamefont {Chen}, \citenamefont {Johansson}, \citenamefont {Roiban},\
  and\ \citenamefont {Zeng}}]{Bern:2017ucb}%
  \BibitemOpen
  \bibfield  {author} {\bibinfo {author} {\bibfnamefont {Z.}~\bibnamefont
  {Bern}}, \bibinfo {author} {\bibfnamefont {J.~J.~M.}\ \bibnamefont
  {Carrasco}}, \bibinfo {author} {\bibfnamefont {W.-M.}\ \bibnamefont {Chen}},
  \bibinfo {author} {\bibfnamefont {H.}~\bibnamefont {Johansson}}, \bibinfo
  {author} {\bibfnamefont {R.}~\bibnamefont {Roiban}},\ and\ \bibinfo {author}
  {\bibfnamefont {M.}~\bibnamefont {Zeng}},\ }\bibfield  {title} {\bibinfo
  {title} {{Five-loop four-point integrand of $N=8$ supergravity as a
  generalized double copy}},\ }\href
  {https://doi.org/10.1103/PhysRevD.96.126012} {\bibfield  {journal} {\bibinfo
  {journal} {Phys. Rev.}\ }\textbf {\bibinfo {volume} {D96}},\ \bibinfo {pages}
  {126012} (\bibinfo {year} {2017}{\natexlab{b}})},\ \Eprint
  {https://arxiv.org/abs/1708.06807} {arXiv:1708.06807 [hep-th]} \BibitemShut
  {NoStop}%
\bibitem [{\citenamefont {Donagi}\ and\ \citenamefont
  {Witten}(2015)}]{Donagi:2013dua}%
  \BibitemOpen
  \bibfield  {author} {\bibinfo {author} {\bibfnamefont {R.}~\bibnamefont
  {Donagi}}\ and\ \bibinfo {author} {\bibfnamefont {E.}~\bibnamefont
  {Witten}},\ }\bibfield  {title} {\bibinfo {title} {{Supermoduli Space Is Not
  Projected}},\ }\href@noop {} {\bibfield  {journal} {\bibinfo  {journal}
  {Proc. Symp. Pure Math.}\ }\textbf {\bibinfo {volume} {90}},\ \bibinfo
  {pages} {19} (\bibinfo {year} {2015})},\ \Eprint
  {https://arxiv.org/abs/1304.7798} {arXiv:1304.7798 [hep-th]} \BibitemShut
  {NoStop}%
\bibitem [{\citenamefont {Berkovits}(2007)}]{Berkovits:2006vc}%
  \BibitemOpen
  \bibfield  {author} {\bibinfo {author} {\bibfnamefont {N.}~\bibnamefont
  {Berkovits}},\ }\bibfield  {title} {\bibinfo {title} {{New higher-derivative
  R**4 theorems}},\ }\href {https://doi.org/10.1103/PhysRevLett.98.211601}
  {\bibfield  {journal} {\bibinfo  {journal} {Phys. Rev. Lett.}\ }\textbf
  {\bibinfo {volume} {98}},\ \bibinfo {pages} {211601} (\bibinfo {year}
  {2007})},\ \Eprint {https://arxiv.org/abs/hep-th/0609006}
  {arXiv:hep-th/0609006} \BibitemShut {NoStop}%
\bibitem [{\citenamefont {He}\ and\ \citenamefont {Yuan}(2015)}]{He:2015yua}%
  \BibitemOpen
  \bibfield  {author} {\bibinfo {author} {\bibfnamefont {S.}~\bibnamefont
  {He}}\ and\ \bibinfo {author} {\bibfnamefont {E.~Y.}\ \bibnamefont {Yuan}},\
  }\bibfield  {title} {\bibinfo {title} {{One-loop Scattering Equations and
  Amplitudes from Forward Limit}},\ }\href
  {https://doi.org/10.1103/PhysRevD.92.105004} {\bibfield  {journal} {\bibinfo
  {journal} {Phys. Rev.}\ }\textbf {\bibinfo {volume} {D92}},\ \bibinfo {pages}
  {105004} (\bibinfo {year} {2015})},\ \Eprint
  {https://arxiv.org/abs/1508.06027} {arXiv:1508.06027 [hep-th]} \BibitemShut
  {NoStop}%
\bibitem [{\citenamefont {Baadsgaard}\ \emph {et~al.}(2016)\citenamefont
  {Baadsgaard}, \citenamefont {Bjerrum-Bohr}, \citenamefont {Bourjaily},
  \citenamefont {Caron-Huot}, \citenamefont {Damgaard},\ and\ \citenamefont
  {Feng}}]{Baadsgaard:2015twa}%
  \BibitemOpen
  \bibfield  {author} {\bibinfo {author} {\bibfnamefont {C.}~\bibnamefont
  {Baadsgaard}}, \bibinfo {author} {\bibfnamefont {N.~E.~J.}\ \bibnamefont
  {Bjerrum-Bohr}}, \bibinfo {author} {\bibfnamefont {J.~L.}\ \bibnamefont
  {Bourjaily}}, \bibinfo {author} {\bibfnamefont {S.}~\bibnamefont
  {Caron-Huot}}, \bibinfo {author} {\bibfnamefont {P.~H.}\ \bibnamefont
  {Damgaard}},\ and\ \bibinfo {author} {\bibfnamefont {B.}~\bibnamefont
  {Feng}},\ }\bibfield  {title} {\bibinfo {title} {{New Representations of the
  Perturbative S-Matrix}},\ }\href
  {https://doi.org/10.1103/PhysRevLett.116.061601} {\bibfield  {journal}
  {\bibinfo  {journal} {Phys. Rev. Lett.}\ }\textbf {\bibinfo {volume} {116}},\
  \bibinfo {pages} {061601} (\bibinfo {year} {2016})},\ \Eprint
  {https://arxiv.org/abs/1509.02169} {arXiv:1509.02169 [hep-th]} \BibitemShut
  {NoStop}%
\bibitem [{\citenamefont {Cachazo}\ \emph {et~al.}(2016)\citenamefont
  {Cachazo}, \citenamefont {He},\ and\ \citenamefont {Yuan}}]{Cachazo:2015aol}%
  \BibitemOpen
  \bibfield  {author} {\bibinfo {author} {\bibfnamefont {F.}~\bibnamefont
  {Cachazo}}, \bibinfo {author} {\bibfnamefont {S.}~\bibnamefont {He}},\ and\
  \bibinfo {author} {\bibfnamefont {E.~Y.}\ \bibnamefont {Yuan}},\ }\bibfield
  {title} {\bibinfo {title} {{One-Loop Corrections from Higher Dimensional Tree
  Amplitudes}},\ }\href {https://doi.org/10.1007/JHEP08(2016)008} {\bibfield
  {journal} {\bibinfo  {journal} {JHEP}\ }\textbf {\bibinfo {volume} {08}},\
  \bibinfo {pages} {008}},\ \Eprint {https://arxiv.org/abs/1512.05001}
  {arXiv:1512.05001 [hep-th]} \BibitemShut {NoStop}%
\bibitem [{\citenamefont {He}\ and\ \citenamefont
  {Schlotterer}(2017)}]{He:2016mzd}%
  \BibitemOpen
  \bibfield  {author} {\bibinfo {author} {\bibfnamefont {S.}~\bibnamefont
  {He}}\ and\ \bibinfo {author} {\bibfnamefont {O.}~\bibnamefont
  {Schlotterer}},\ }\bibfield  {title} {\bibinfo {title} {{New Relations for
  Gauge-Theory and Gravity Amplitudes at Loop Level}},\ }\href
  {https://doi.org/10.1103/PhysRevLett.118.161601} {\bibfield  {journal}
  {\bibinfo  {journal} {Phys. Rev. Lett.}\ }\textbf {\bibinfo {volume} {118}},\
  \bibinfo {pages} {161601} (\bibinfo {year} {2017})},\ \Eprint
  {https://arxiv.org/abs/1612.00417} {arXiv:1612.00417 [hep-th]} \BibitemShut
  {NoStop}%
\bibitem [{\citenamefont {Feng}(2016)}]{Feng:2016nrf}%
  \BibitemOpen
  \bibfield  {author} {\bibinfo {author} {\bibfnamefont {B.}~\bibnamefont
  {Feng}},\ }\bibfield  {title} {\bibinfo {title} {{CHY-construction of Planar
  Loop Integrands of Cubic Scalar Theory}},\ }\href
  {https://doi.org/10.1007/JHEP05(2016)061} {\bibfield  {journal} {\bibinfo
  {journal} {JHEP}\ }\textbf {\bibinfo {volume} {05}},\ \bibinfo {pages}
  {061}},\ \Eprint {https://arxiv.org/abs/1601.05864} {arXiv:1601.05864
  [hep-th]} \BibitemShut {NoStop}%
\bibitem [{\citenamefont {He}\ \emph {et~al.}(2018)\citenamefont {He},
  \citenamefont {Schlotterer},\ and\ \citenamefont {Zhang}}]{He:2017spx}%
  \BibitemOpen
  \bibfield  {author} {\bibinfo {author} {\bibfnamefont {S.}~\bibnamefont
  {He}}, \bibinfo {author} {\bibfnamefont {O.}~\bibnamefont {Schlotterer}},\
  and\ \bibinfo {author} {\bibfnamefont {Y.}~\bibnamefont {Zhang}},\ }\bibfield
   {title} {\bibinfo {title} {{New BCJ representations for one-loop amplitudes
  in gauge theories and gravity}},\ }\href
  {https://doi.org/10.1016/j.nuclphysb.2018.03.003} {\bibfield  {journal}
  {\bibinfo  {journal} {Nucl. Phys.}\ }\textbf {\bibinfo {volume} {B930}},\
  \bibinfo {pages} {328} (\bibinfo {year} {2018})},\ \Eprint
  {https://arxiv.org/abs/1706.00640} {arXiv:1706.00640 [hep-th]} \BibitemShut
  {NoStop}%
\bibitem [{\citenamefont {Geyer}\ and\ \citenamefont
  {Monteiro}(2018{\natexlab{b}})}]{Geyer:2017ela}%
  \BibitemOpen
  \bibfield  {author} {\bibinfo {author} {\bibfnamefont {Y.}~\bibnamefont
  {Geyer}}\ and\ \bibinfo {author} {\bibfnamefont {R.}~\bibnamefont
  {Monteiro}},\ }\bibfield  {title} {\bibinfo {title} {{Gluons and gravitons at
  one loop from ambitwistor strings}},\ }\href
  {https://doi.org/10.1007/JHEP03(2018)068} {\bibfield  {journal} {\bibinfo
  {journal} {JHEP}\ }\textbf {\bibinfo {volume} {03}},\ \bibinfo {pages}
  {068}},\ \Eprint {https://arxiv.org/abs/1711.09923} {arXiv:1711.09923
  [hep-th]} \BibitemShut {NoStop}%
\bibitem [{\citenamefont {Geyer}\ \emph {et~al.}(2019)\citenamefont {Geyer},
  \citenamefont {Monteiro},\ and\ \citenamefont
  {Stark-Muchão}}]{Geyer:2019hnn}%
  \BibitemOpen
  \bibfield  {author} {\bibinfo {author} {\bibfnamefont {Y.}~\bibnamefont
  {Geyer}}, \bibinfo {author} {\bibfnamefont {R.}~\bibnamefont {Monteiro}},\
  and\ \bibinfo {author} {\bibfnamefont {R.}~\bibnamefont {Stark-Muchão}},\
  }\bibfield  {title} {\bibinfo {title} {{Two-Loop Scattering Amplitudes:
  Double-Forward Limit and Colour-Kinematics Duality}},\ }\href
  {https://doi.org/10.1007/JHEP12(2019)049} {\bibfield  {journal} {\bibinfo
  {journal} {JHEP}\ }\textbf {\bibinfo {volume} {12}},\ \bibinfo {pages}
  {049}},\ \Eprint {https://arxiv.org/abs/1908.05221} {arXiv:1908.05221
  [hep-th]} \BibitemShut {NoStop}%
\bibitem [{\citenamefont {Edison}\ \emph {et~al.}(2020)\citenamefont {Edison},
  \citenamefont {He}, \citenamefont {Schlotterer},\ and\ \citenamefont
  {Teng}}]{Edison:2020uzf}%
  \BibitemOpen
  \bibfield  {author} {\bibinfo {author} {\bibfnamefont {A.}~\bibnamefont
  {Edison}}, \bibinfo {author} {\bibfnamefont {S.}~\bibnamefont {He}}, \bibinfo
  {author} {\bibfnamefont {O.}~\bibnamefont {Schlotterer}},\ and\ \bibinfo
  {author} {\bibfnamefont {F.}~\bibnamefont {Teng}},\ }\bibfield  {title}
  {\bibinfo {title} {{One-loop Correlators and BCJ Numerators from Forward
  Limits}},\ }\href {https://doi.org/10.1007/JHEP09(2020)079} {\bibfield
  {journal} {\bibinfo  {journal} {JHEP}\ }\textbf {\bibinfo {volume} {09}},\
  \bibinfo {pages} {079}},\ \Eprint {https://arxiv.org/abs/2005.03639}
  {arXiv:2005.03639 [hep-th]} \BibitemShut {NoStop}%
\bibitem [{\citenamefont {Farrow}\ \emph {et~al.}(2020)\citenamefont {Farrow},
  \citenamefont {Geyer}, \citenamefont {Lipstein}, \citenamefont {Monteiro},\
  and\ \citenamefont {Stark-Much\~ao}}]{Farrow:2020voh}%
  \BibitemOpen
  \bibfield  {author} {\bibinfo {author} {\bibfnamefont {J.~A.}\ \bibnamefont
  {Farrow}}, \bibinfo {author} {\bibfnamefont {Y.}~\bibnamefont {Geyer}},
  \bibinfo {author} {\bibfnamefont {A.~E.}\ \bibnamefont {Lipstein}}, \bibinfo
  {author} {\bibfnamefont {R.}~\bibnamefont {Monteiro}},\ and\ \bibinfo
  {author} {\bibfnamefont {R.}~\bibnamefont {Stark-Much\~ao}},\ }\bibfield
  {title} {\bibinfo {title} {{Propagators, BCFW recursion and new scattering
  equations at one loop}},\ }\href {https://doi.org/10.1007/JHEP10(2020)074}
  {\bibfield  {journal} {\bibinfo  {journal} {JHEP}\ }\textbf {\bibinfo
  {volume} {10}},\ \bibinfo {pages} {074}},\ \Eprint
  {https://arxiv.org/abs/2007.00623} {arXiv:2007.00623 [hep-th]} \BibitemShut
  {NoStop}%
\bibitem [{\citenamefont {Mafra}\ \emph {et~al.}(2013)\citenamefont {Mafra},
  \citenamefont {Schlotterer},\ and\ \citenamefont
  {Stieberger}}]{Mafra:2011nv}%
  \BibitemOpen
  \bibfield  {author} {\bibinfo {author} {\bibfnamefont {C.~R.}\ \bibnamefont
  {Mafra}}, \bibinfo {author} {\bibfnamefont {O.}~\bibnamefont {Schlotterer}},\
  and\ \bibinfo {author} {\bibfnamefont {S.}~\bibnamefont {Stieberger}},\
  }\bibfield  {title} {\bibinfo {title} {{Complete N-Point Superstring Disk
  Amplitude I. Pure Spinor Computation}},\ }\href
  {https://doi.org/10.1016/j.nuclphysb.2013.04.023} {\bibfield  {journal}
  {\bibinfo  {journal} {Nucl. Phys.}\ }\textbf {\bibinfo {volume} {B873}},\
  \bibinfo {pages} {419} (\bibinfo {year} {2013})},\ \Eprint
  {https://arxiv.org/abs/1106.2645} {arXiv:1106.2645 [hep-th]} \BibitemShut
  {NoStop}%
\bibitem [{\citenamefont {Azevedo}\ \emph {et~al.}(2018)\citenamefont
  {Azevedo}, \citenamefont {Chiodaroli}, \citenamefont {Johansson},\ and\
  \citenamefont {Schlotterer}}]{Azevedo:2018dgo}%
  \BibitemOpen
  \bibfield  {author} {\bibinfo {author} {\bibfnamefont {T.}~\bibnamefont
  {Azevedo}}, \bibinfo {author} {\bibfnamefont {M.}~\bibnamefont {Chiodaroli}},
  \bibinfo {author} {\bibfnamefont {H.}~\bibnamefont {Johansson}},\ and\
  \bibinfo {author} {\bibfnamefont {O.}~\bibnamefont {Schlotterer}},\
  }\bibfield  {title} {\bibinfo {title} {{Heterotic and bosonic string
  amplitudes via field theory}},\ }\href
  {https://doi.org/10.1007/JHEP10(2018)012} {\bibfield  {journal} {\bibinfo
  {journal} {JHEP}\ }\textbf {\bibinfo {volume} {10}},\ \bibinfo {pages}
  {012}},\ \Eprint {https://arxiv.org/abs/1803.05452} {arXiv:1803.05452
  [hep-th]} \BibitemShut {NoStop}%
\bibitem [{\citenamefont {He}\ \emph {et~al.}(2019)\citenamefont {He},
  \citenamefont {Teng},\ and\ \citenamefont {Zhang}}]{He:2018pol}%
  \BibitemOpen
  \bibfield  {author} {\bibinfo {author} {\bibfnamefont {S.}~\bibnamefont
  {He}}, \bibinfo {author} {\bibfnamefont {F.}~\bibnamefont {Teng}},\ and\
  \bibinfo {author} {\bibfnamefont {Y.}~\bibnamefont {Zhang}},\ }\bibfield
  {title} {\bibinfo {title} {{String amplitudes from field-theory amplitudes
  and vice versa}},\ }\href {https://doi.org/10.1103/PhysRevLett.122.211603}
  {\bibfield  {journal} {\bibinfo  {journal} {Phys. Rev. Lett.}\ }\textbf
  {\bibinfo {volume} {122}},\ \bibinfo {pages} {211603} (\bibinfo {year}
  {2019})},\ \Eprint {https://arxiv.org/abs/1812.03369} {arXiv:1812.03369
  [hep-th]} \BibitemShut {NoStop}%
\bibitem [{\citenamefont {Igusa}(1967)}]{Igusa}%
  \BibitemOpen
  \bibfield  {author} {\bibinfo {author} {\bibfnamefont {J.}~\bibnamefont
  {Igusa}},\ }\bibfield  {title} {\bibinfo {title} {{Modular Forms and
  Projective Invariants}},\ }\href {https://doi.org/10.2307/2373243} {\bibfield
   {journal} {\bibinfo  {journal} {Amer. J. Math.}\ }\textbf {\bibinfo {volume}
  {89}},\ \bibinfo {pages} {817} (\bibinfo {year} {1967})}\BibitemShut
  {NoStop}%
\end{thebibliography}%

\end{document}